\documentclass[trackchanges,twocolumn]{aastex63}
\usepackage{multirow}

\usepackage{orcidlink}
\usepackage{amsmath}

\begin{document}

\title{Detecting electromagnetic counterparts to LIGO/Virgo/KAGRA gravitational wave events with DECam: Neutron Star Mergers}

\author[0009-0000-4830-1484]{K.~Kunnumkai}
\author[0000-0002-6011-0530]{A.~Palmese}
\affiliation{McWilliams Center for Cosmology and Astrophysics, Department of Physics, Carnegie Mellon University, Pittsburgh, PA 15213, USA}
\author[0000-0002-6121-0285]{A. M. Farah}
\affiliation{Department of Physics, University of Chicago, Chicago, IL 60637, USA}
\author[0000-0002-8255-5127]{M.~Bulla}
\affiliation{Department of Physics and Earth Science, University of Ferrara, via Saragat 1, I-44122 Ferrara, Italy}
\affiliation{INFN, Sezione di Ferrara, via Saragat 1, I-44122 Ferrara, Italy}
\affiliation{INAF, Osservatorio Astronomico d’Abruzzo, via Mentore Maggini snc, 64100 Teramo, Italy}
\author[0000-0003-2374-307X]{T.~Dietrich}
\affiliation{Institut f\"{u}r Physik und Astronomie, Universit\"{a}t Potsdam, Haus 28, Karl-Liebknecht-Str. 24/25, 14476, Potsdam, Germany}
\affiliation{Max Planck Institute for Gravitational Physics (Albert Einstein Institute), Am M\"{u}hlenberg 1, Potsdam 14476, Germany}
\author[0000-0001-7041-3239]{P.~T.~H.~Pang}
\affiliation{Institute for Gravitational and Subatomic Physics (GRASP), Utrecht University, Princetonplein 1, 3584 CC Utrecht, The Netherlands}
\affiliation{Nikhef, Science Park 105, 1098 XG Amsterdam, The Netherlands}
\author[0000-0003-3768-7515]{S. Anand}
\affiliation{Cahill Center for Astrophysics, California Institute of Technology, Pasadena CA 91125, USA}
\author[0000-0002-8977-1498]{I. Andreoni}\affiliation{Department of Physics and Astronomy, University of North Carolina at Chapel Hill, Chapel Hill, NC 27599-3255, USA}
\author[0000-0002-1270-7666]{T. Cabrera}\affiliation{McWilliams Center for Cosmology and Astrophysics, Department of Physics, Carnegie Mellon University, Pittsburgh, PA 15213, USA}
\author[0000-0002-9700-0036]{B. O'Connor}
\affiliation{McWilliams Center for Cosmology and Astrophysics, Department of Physics, Carnegie Mellon University, Pittsburgh, PA 15213, USA} 

\begin{abstract}

With GW170817 being the only multimessenger gravitational wave (GW) event with an associated kilonova detected so far, there exists a pressing need for realistic estimation of the GW localization uncertainties and rates, as well as optimization of available telescope time to enable the detection of new kilonovae. We simulate GW events assuming a data-driven distribution of binary parameters for the LIGO/Virgo/KAGRA (LVK) fourth and fifth observing runs (O4 and O5). We map the binary neutron star (BNS) and neutron star-black hole (NSBH) properties to the kilonova optical light curves. We use the simulated population of kilonovae to generate follow-up observing plans, with the primary goal of optimizing detection with the Gravitational Wave Multi-Messenger Astronomy DECam Survey (GW-MMADS). We explore the dependence of kilonova detectability on the mass, distance, inclination, and spin of the binaries. Assuming that no BNS was detected during O4 until the end of 2024, we present updated GW BNS (NSBH) merger detection rates. We expect to detect BNS (NSBH) kilonovae with DECam at a per-year rate of: $0$--$2.0$ ($0$) in O4, and $2.0$--$19$ ($0$--$1.0$) in O5. We expect the majority of BNS detections and also those accompanied by a detectable kilonova to produce a hypermassive NS remnant, with a significant fraction of the remaining BNSs promptly collapsing to a BH. We release GW simulations and depths required to detect kilonovae based on our predictions to support the astronomical community in their multimessenger follow-up campaigns and analyses.
\end{abstract}


\section{Introduction}

In 2017, a binary neutron star (BNS) merger event was detected in gravitational waves (GWs), GW170817 \citep{ligobns}, and it was accompanied by multiple electromagnetic (EM) counterparts \citep{MMApaper}. The detection of its EM counterparts, namely the gamma-ray burst GRB 170817A \citep{LIGOScientific:2017zic,2017GCN.21528....1G,Savchenko_2017,Pozanenko_2018}, the kilonova (KN) AT2017gfo \citep{Coulter1556,Soares_Santos_2017,arcavi,2017ApJ...848L..17C,HU20171433,Smartt_2017,tanvir}, and later the afterglow \citep{Fong2019,Makhathini_2021}, opened a new era of multimessenger astronomy \citep{margutti_review}. 

Kilonovae \citep{Rosswog_2005,2013ApJ...775...18B,2014MNRAS.439..757G,metzger_kilonovae} are transients ranging from the infrared to the ultraviolet powered by the radioactive decay of heavy nuclei synthesized in the highly-neutronized material ejected during neutron star (NS) mergers. These transients are considered to provide smoking-gun evidence for BNS or neutron star-black hole (NSBH) mergers. The detection of the KN AT2017gfo enabled hundreds of multimessenger analyses on the theory of gravity \citep{Belgacem_2018}, the Universe’s expansion \citep{2017Natur.551...85A,Hotokezaka19,2022Univ....8..289B,Palmese_2023,Palmese:2023beh}, the formation and fate of compact objects \citep{Blanchard_2017,palmese,tsai2020gw170817,Kilpatrick_2022}, thanks to its precise localization in the sky, which allowed us to pinpoint the host galaxy of the GW merger \citep{Coulter1556,Blanchard_2017,Hjorth_2017,Abbott_2017,palmese}. Moreover, KN observations themselves provide us with information on the physics of NS mergers, including but not limited to their composition \citep{2019Natur.574..497W,2021ApJ...913...26D,anand2023chemical,2023MNRAS.526L.155H,2023ApJ...944..123V,2023A&A...675A.194S}, EOS \citep{Coughlin_2018,Margalit_2019,Dietrich_2020,P_rez_Garc_a_2022}, fate of the remnant object \citep{Yu_2018} and origin of heavy elements (generated by r-process nucleosynthesis in the neutron-rich ejecta of the KNe \citep{Kasen_2017,Pian_2017}). The wide variety of possible multimessenger analyses from the identification of a KN motivates us to optimize the follow-up of BNS and NSBH merger events to characterize multimessenger sources beyond GW170817.

Currently, the international GW network consisting of the advanced LIGO~\citep{2018LRR....21....3A}, advanced Virgo~\citep{VIRGO:2014yos}, and KAGRA~\citep{KAGRA:2020tym} interferometers (LVK) is carrying out its fourth observing run (O4). Although the LVK detected a few BNS \citep{Abbott_2020_GW190425} and NSBH \citep{Abbott_2021,Abbott_2023} candidates at different confidence levels in O3, there has been no clear multi-messenger KN detection since the second observing run (O2) \citep{2019ApJ...884L..55G,Anand_2020,Andreoni_2020}. This is despite the improved detector sensitivity beyond O2, which should lead to a larger number of NS merger detections with an increased signal-to-noise ratio. Despite the fact that EM counterparts other than KNe have been claimed to be associated with GW events \citep{Graham_2020,2023NatAs...7..579M}, their association is uncertain \citep{ashton,Palmese_2021,bhardwaj2023challenges,smartt2023gw190425,Radice_2024,magana24}. There have also been no multimessenger KN detection in O4 so far despite the follow-up efforts \citep{Ahumada_2024,Pillas2025,Hu_2025}. The lack of KN detections can mainly be attributed to both NS merger rates likely to be at the lower end of the 90\% credible interval range previously estimated in  \citet{abbott_population_2023}, the lack of Virgo operations during the first part of O4, and, related to this point, the greater uncertainty in the sky localization of NS mergers from GWs than what was predicted in forecasts including Virgo and often KAGRA (see, e.g., \citealt{Petrov_2022,colombo22,shah24}). This calls for a more realistic estimation of the localization uncertainties of NS mergers and a revised observing strategy to follow up events. To support such efforts, observing scenarios are simulated to accurately predict the source localization and detection of GW events. 

In this work, we make end-to-end simulations of GW events during the current O4 run and the upcoming O5 run (expected to start in 2028) using updated LIGO$/$Virgo$/$KAGRA (LVK) detector sensitivities to assess the ability of the network to reconstruct the sky localization of the source and to accurately forecast the detection rates of EM counterparts. Moreover, we make use of the latest data-driven population for the GW source properties to inject mergers into our simulations \citep{fishbach_does_2020,Farah_2022,abbott_population_2023}. This ensures that the important features in the mass and spin distributions of black holes (BHs) and NSs are properly incorporated in the simulations, since these can have a significant effect on the EM counterpart detectability. First, since the astrophysical GW population of NSs in BNS and NSBH mergers possibly shows a broader and flatter mass distribution than observed from EM observations in the Milky Way \citep{abbott_population_2023}, and because the mass of the binary components has a strong impact on the KN ejecta properties, taking into account the GW population to the best of our knowledge is essential to optimize our follow-up strategy of both BNS and NSBH mergers. Second, the mass distributions considered in \citet{fishbach_does_2020,Farah_2022,abbott_population_2023,Ray_2023} allow for a dip between the expected NS and BH mass distributions, which is not necessarily an empty gap, hence it can accommodate objects such as those in GW190814 \citep{GW190814}. A population model that takes into account objects in the gap may result in a more promising scenario for multi-messenger kilonova detections from NSBHs \citep{230529_LVK,2024arXiv240910651K} than was previously thought \citep{Biscoveanu_2022}. We then compute KN light curves for each modeled BNS and NSBH mergers and develop follow-up strategies based on the KN magnitude distributions.

The full set of GW simulations we present in this work, including the injection parameters, the detection statistics, and the sky maps, is publicly available via Zenodo\footnote{\href{https://doi.org/10.5281/zenodo.14207687}{https://doi.org/10.5281/zenodo.14207687}} \citep{zenodo}. These simulations, discussed in detail in Section \ref{method_GW_sim} and combined with predictions presented in this work, are intended to aid the astronomical community to accurately plan observation programs to minimize wastage of precious telescope time. Specifically, we consider the case of optimizing EM counterpart searches for one of the most powerful optical-to-near-infrared instruments currently available to follow up GW events, the Dark Energy Camera (DECam; \citealt{flaugher}). This is designed for our survey GW-MMADS (Gravitational Wave Multi-Messenger Astronomy DECam Survey; Proposal ID: 2023B-851374; PIs: Andreoni \& Palmese, see e.g. \citealt{cabrera2024}) which is carrying out follow up of GW events during O4. This is an effort to complement previous works \citep{2018LRR....21....3A,Petrov_2022,colombo22,Kiendrebeogo_2023,shah24,bom24,kaur2024} in various ways, namely by taking into account the reduced Virgo sensitivity compared to what was previously considered for O4, by considering a specific follow-up campaign, accounting for uncertainties in the inference of the mass distribution of compact binary mergers, and employing different KN models. 

This paper is structured as follows: in Section \ref{section:Methodology}, we describe the methods used to produce the simulations for the O4 and O5 runs, to attach simulated KNe lightcurves to the GW events and to schedule simulated follow-up campaigns of the events. In Section \ref{section:Results}, we discuss the results of our simulation runs as well as the KN detection efficiency with our fiducial strategy, and its dependence on various binary parameters. We present our discussion and conclusion in Section \ref{section:Discussion} and \ref{section:Conclusion} respectively.

\section{Methodology}\label{section:Methodology}

\subsection{Simulation of GW Signals}
\label{method_GW_sim}
We produce realistic simulations of GW events with LVK O4 and O5 sensitivities. We use BAYESTAR, a rapid sky localization code to localize the mergers in sky and distance \citep{bayestar,Singer_2016}, tools from LALSuite \citep{lalsuite} to make simulations of GW events and the public software \texttt{ligo.skymap} to generate and visualize the GW skymaps. We make 3 different simulations. Both simulation~1 and simulation~2 assume LIGO detectors at O4 sensitivity, but the latter assumes Virgo at O3 sensitivity in addition to the LIGO detectors. For simulation~3, we use all LVK detectors at the expected O5 sensitivity.

\begin{table}[htpb!]
\centering
\begin{tabular}{c c  c  c  c c}  
 \hline
 Simulation & Observing & Network &  \textbf{BNS} & 
 \textbf{NSBH} \\
 number & run\\
 \hline\hline
 1 & O4 & HL  & 720  & 1090\\ 
 \hline
 2 & O4 & HLV & 900  & 1320\\
 \hline
 3 & O5 & HLVK  &  2400 & 4160\\
 \hline \hline
\end{tabular}

\caption{Number of detected BNS and NSBH mergers from different simulations out of the $10^{5}$ compact binary coalescence (CBC) drawn from the distribution of \citealt{Farah_2022}. For simulation 1, we used the LIGO detectors at O4 sensitivity, simulation 2 consists of the Virgo detector at O3 sensitivity in addition to the LIGO detectors at O4 sensitivity, whereas simulation 3 has all LVK detectors at O5 sensitivity. H, L, V, and K stands for LIGO Hanford, LIGO Livingston, Virgo, and KAGRA detectors, respectively.}\label{tab:events}
\end{table}

The mass and spin distributions are chosen as described in the \textsc{Power Law + Dip + Break} (PDB) model \citep{fishbach_does_2020,Farah_2022,abbott_population_2023}. It assumes a broken power law in mass with a notch filter to enable a mass gap between $\sim2 - 8 ~M_{\odot}$ that is allowed to vary in depth. The model also adds a low-pass filter to the upper end of the BH mass to allow for possible tapering of the mass distribution. The masses of both components (i.e. primary and secondary mass, $m_1$ and $m_2$) are described by this functional form, and a pairing function is employed to account for the fact that, within a given binary, the components' masses tend to be similar.

Draws are taken from the hyper-posterior of the PDB model fit to the third GW Transient Catalog (GWTC-3; \citealt{gwtc3}). These hyper-posterior draws describe the probability of obtaining a system with masses $m_1$ and $m_2$: $p(m_1,m_2)$. The masses are then drawn from $p(m_1,m_2)$. Note that this differs from the population assumed in \citealt{Kiendrebeogo_2023}, although that work also employed the PDB model: here we take into account the uncertainties in the population inference as we consider the full population posterior, effectively marginalizing over these uncertainties, instead of assuming a fixed population from the maximum \emph{a posteriori} parameters. This is especially important for EM-bright NSBHs as the PDB posterior mode from GWTC-3 happens to have a nearly empty mass gap, but the uncertainties on the depth and location of the gap are large, so that considering the entire range of possibilities allowed by the data may have a significant impact. 

Our model additionally assumes an isotropic orientation and a uniform magnitude for component spins. Objects with a mass less than 2.5 $M_{\odot}$ have spin magnitudes in the range $[0,0.4]$, while objects with higher masses have spins defined as in the range $[0,1]$. As described in \citealt{abbott_population_2023}, these spin ranges were established by the sensitivity estimates available in GWTC-3.

We draw $10^{5}$ CBCs from a sample following the PDB model. We draw the masses and spins of the CBCs from the distribution described above, uniformly distribute the samples in comoving volume, and isotropically distribute them in the orbital orientation and sky location. We assume an isotropic distribution of angular momentum directions. We follow \citealt{Petrov_2022} and assess the detectability of GW events with an individual detector and network signal-to-noise ratio (SNR) thresholds of 1 and 8, respectively, for the population of NS mergers. If at least one of the detectors detects the signal above these thresholds, it is counted as a GW event. For all of the simulations discussed in this work, we use a duty cycle of 70$\%$ for each detector and employ a Gaussian filter to recover the signal from the noise. The statistics of the sources detected, grouped by subpopulation, are shown in Table \ref{tab:events}. The noise power spectral density (PSD) curves are considered for each detector per observing run. For the LIGO network at O4 sensitivity, we use the PSD curve aligo\_O4high.txt \citep{Abbott_2020}, and for the Virgo detector at O3 sensitivity, we use avirgo\_O3actual.txt \citep{Abbott_2020}. For the O5 run, we assume an A+ sensitivity for the LIGO detectors, avirgo\_O5low\_NEW.txt for Virgo and kagra\_128Mpc.txt for KAGRA respectively, all from \url{https://dcc.ligo.org/LIGO-T2000012-v1/public}. 

The distribution we use does not rely on binary classification, allowing us to define mass ranges for the three astrophysical sub-populations in which we are interested: BNSs, NSBHs, and binary black holes (BBHs). The maximum mass of a non-rotating NS $M_{\rm TOV}$ is calculated using the Tolman-Oppenheimer-Volkoff (TOV) limit \citep{Kalogera_1996}, which according to the maximum posterior EOS from \citealt{Huth:2021bsp} assumed in this work is 2.436 $M_{\odot}$. A rotating NS can exceed this limit. To account for this, we compute the maximum mass the components can have in each BNS and NSBH systems based on the spin $a$ of each component, using the equation given in \citealt{Breu_2016,2020MNRAS.499L..82M}:
\begin{equation}
    M_{\rm max} = M_{\rm TOV}\left(1+A_{2}\left(\frac{a}{a_{\rm Kep}}\right)^{2} + A_{4}\left(\frac{a}{a_{\rm Kep}}\right)^{4}\right)
    \label{eq:Mmax}
\end{equation}
where $A_{\rm 2}$ = 0.132, $A_{\rm 4}$ = 0.071 and for our fiducial EOS the dimensionless spin at the mass shedding limit is $a_{\rm Kep}$ = 0.606, and it is defined in \citealt{2020MNRAS.499L..82M,Breu_2016}:
\begin{equation}
    a_{\rm Kep} = \frac{\alpha_{\rm 1}}{\sqrt{C_{\rm TOV}}} + \alpha_{\rm 2}\sqrt{\rm C_{\rm TOV}}
\end{equation} 
Here $\alpha_{\rm 1}$ = 0.045, $\alpha_{\rm 2}$ = 1.112 and $C_{\rm TOV}$ is related to the TOV mass and radius by the equation $C_{\rm TOV}$ = $\frac{M_{\rm TOV}}{R_{\rm TOV}}$.
For the purposes of classification, we use the maximum NS mass $M_{\rm max}$ defined in Eq. (\ref{eq:Mmax}). Labeling the more massive object in each system as $m_1$ and the less massive object as $m_2$, the BNS systems are then binaries with $m_1,m_2<M_{\rm max}$, NSBH are those with $m_2<M_{\rm max}$ and $m_1\geq M_{\rm max}$. 

In what follows, we also explore the detectability of KN assuming the lower and upper 95\% credible intervals of the EOS constraints from \citealt{Huth:2021bsp}, hereafter referred to as softer and stiffer EOS, to assess the impact of the EOS on our results. We define “softer”, “fiducial”, and “stiffer” EOSs based on the 2.5\%, 50\%, and 97.5\% quantiles of the marginalized posterior distribution of the tidal deformability of a 1.4 $M_{\odot}$ neutron star ($\lambda_{1.4}$). This distribution is derived from the EOS ensemble released by \citet{Huth:2021bsp} where each EOS is associated with a probability weight based on its astrophysics and nuclear physics-based constraints. For each EOS in the ensemble, the $\lambda_{1.4}$ value is computed by interpolating the relation between mass and tidal deformability. The resulting list of $\lambda_{1.4}$ values is plotted as a histogram using the posterior weights provided at Zenodo\footnote{\href{https://zenodo.org/records/6106130}{https://zenodo.org/records/6106130}} \citep{zenodo_lamda}. The EOSs at the lower and upper 2.5\% percentiles of this weighted distribution are designated as “soft” and “stiff” respectively. $M_{\rm TOV}$ for softer, fiducial and stiffer EOSs used in this work are 2.069 $M_{\odot}$, 2.436 $M_{\odot}$ and 2.641 $M_{\odot}$ respectively.

\subsection{Simulation of EM Counterparts}
\label{section:Methods-counterpart}

For modeling KN light curves from BNS and NSBH mergers, we use the Bu2019lm and Bu2019nsbh surrogate models implemented in the Nuclear physics and MultiMessenger Astronomy (\texttt{NMMA}; \citealt{Dietrich_2020,Pang_2023}) framework. These models are constructed from KN grids computed with the time-dependent three-dimensional Monte Carlo radiative transfer code \texttt{POSSIS} \citep{Bulla_2019} for BNS \citep{Dietrich_2020} and NSBH \citep{Anand_2020} mergers, and allow us to simulate KN light curves for arbitrary combinations of the BNS and NSBH binary parameters. 

The main outflow properties to model a KN with the aforementioned models are the dynamical ejecta mass ($M_{\rm dyn}$) and the disk wind ejecta mass ($M_{\rm wind}$). Dynamical ejecta are launched during the merger due to hydrodynamical forces that squeeze the material at the interface between the merging objects \citep{Bauswein_2013, Hotokezaka_2013}, while the wind ejecta originate from the disk of material accreted around the central remnant. The relative importance of these processes depends on the parameters of the binary such as the chirp mass, mass ratio ($q = m_{2}/m_{1}$) and the EOS, with asymmetric mass-ratio systems ($q \ll 1$) typically leaving significantly larger amounts of unbound material than more symmetric mass-ratio binaries in BNS systems \citep{Bauswein_2013}. 

For BNS mergers, $M_{\rm dyn}$ depends on the compactness of the merging stars and their mass ratio. We follow the relation from \citealt{Kr_ger_2020} to model $M_{\rm dyn}$
\begin{equation}
\label{eq:mdyn_bns}
    \frac{M_{\rm dyn fit}^{\rm BNS}}{10^{-3}} = \left (\frac{a}{C_1} + b \left (\frac{m_2}{m_1} \right)^{n} +cC_{1}\right)m_{1} + (1 \leftrightarrow 2).
\end{equation}
Here, $C_{1} = Gm_{1}/R_{1}c^{2}$ and the best-fit parameters from \citealt{Kr_ger_2020} are $a= -9.3335$, $b = 114.17$, $c = -337.56$ and $n = 1.5465$. The fitting formula in eq. (\ref{eq:mdyn_bns}) is calibrated against numerical simulations and exhibits residuals that are modeled as a Gaussian distribution centered at 0 with standard deviation $\sigma = 0.004 M_{\odot}$ \citep{Kr_ger_2020}.  We call $\alpha$ a random variable drawn from this distribution to capture model uncertainty. Thus the dynamical ejecta mass is modeled as:
\begin{equation}
\label{eq:alphadyn_bns}
    M_{\rm dyn}^{\rm BNS} = M_{\rm dyn fit}^{\rm BNS} + \alpha
\end{equation} 
where $\alpha \sim \mathcal{N}$ (0, 0.004 $M_{\odot}$). $M_{\rm dyn}^{\rm BNS}$ is truncated at zero (no negative values are allowed). Such uncertainties in model arise from the limited coverage of the parameter space by existing numerical simulations, and also from ignoring or approximating some important physics \citep{Kr_ger_2020}, e.g., neutrino radiation.

The disk mass for BNS mergers depends on the total mass and the threshold mass $M_{\rm th}$ (which is the limiting total binary mass $M_{\rm tot}$ beyond which the BNS system would undergo prompt collapse into a BH; ~\citealt{Bauswein:2013jpa,Agathos:2019sah}), and follows the relation from \citealt{Dietrich_2020}:
\begin{equation}
    \label{eq:mdisk_bns}
    \log_{\rm 10}\left(M_{\rm disk}^{\rm BNS}\right) = \max \left (-3, a\left(1 + b \tanh \left [\frac{c-{M_{\rm tot}}/{M_{\rm th}}}{d} \right] \right) \right)\, .
\end{equation}

Here, $M_{\rm th} = k_{\rm th}M_{\rm TOV}$ \citep{Hotokezaka_2011} where $k_{\rm th}$ is a function of $M_{\rm TOV}$ and the EOS \citep{Bauswein:2013jpa}, $a$ and $b$ are functions of the mass ratio $q$ and given by:
\begin{equation}
  \begin{split}
    &a =a_{0} + \delta_{a}x_{i} \\
    &b =  b_{0} + \delta_{b}x_{i}\\
    &x_i = 0.5 \tanh(\beta(q - q_{t})) \\
  \end{split}
\end{equation}
where $a_{0} = -1.581$, $b_{0} = -0.538$, $c = 0.953$, $d = 0.0417$, $\delta_{a}=-2.439$, $\delta_{b} = -0.406$, $\beta=3.910$, and $q_{t}=0.900$. The fraction of disk mass ejected as wind is given by the parameter $\zeta$, which, when multiplied by $M_{\rm disk}^{\rm BNS}$, gives $M_{\rm wind}^{\rm BNS}$. Although the true value of $\zeta$ is unknown \citep{2020PhR...886....1N}, we draw from a Gaussian centered at 0.3 with a standard deviation $\sigma$ of 0.15; cf.~Sec.~2.2 of~\citealt{Raaijmakers:2021slr} for a more detailed discussion.

For NSBH mergers, $M_{\rm dyn}$ is a function of the mass of the components, their spins, the baryonic mass of NS $M_{\rm NS}^{b}$, and the compactness of the secondary component. It is given by the equation from \citealt{Kr_ger_2020}:
\begin{equation}
    \label{eq:M_dyn_nsbh}
    \frac{M_{\rm dyn}^{\rm NSBH}}{M_{\rm NS}^{b}} = a_{1}Q^{n_{1}}\frac{1-2C_{\rm NS}}{C_{\rm NS}} - a_{2}Q^{n_{2}}\frac{R_{\rm ISCO}}{m_1} + a_{4},
\end{equation}
where the best fitting parameters are $a_1 = 0.007116$, $a_2 = 0.001436$, $a_4 =-0.02762$, $n_1 = 0.8636$, and $n_2 = 1.6840$. Here, $Q=m_1/m_2$ and $R_{\rm ISCO}$ is the radius of the innermost stable circular orbit (ISCO) of the BH with mass $m_1$ and spin $\chi_{\rm BH}$. The baryonic mass of the NS is given by $M_{\rm NS}^{b}$ = $m_2 \left(1 + \frac{0.6C_{\rm NS}}{1-0.5C_{\rm NS}}\right)$ \citep{Lattimer_2001}. Again, negative values of $M_{\rm dyn}^{\rm NSBH}$ are assumed to be zero. In our simulations, we add the uncertainty on the dynamical ejecta mass $\alpha$, which is a truncated Gaussian with $\mu = 0$ and $\sigma=0.0047$ \citep{Kr_ger_2020}. 

For NSBH mergers, the remnant mass, which is used to refer to the baryon mass outside the BH $\sim 10~\mathrm{ms}$ post-merger, is calculated using the component masses, spins, and compactness of the NS from the fitting formula by \citealt{Foucart_2018}:
\begin{equation}
    \label{eq:mdisk_nsbh}
    \hat{M}^{\rm NSBH}_{\rm rem} = \left [{\rm max}\left( a\frac{1-2C_{\rm NS}}{\eta^{1/3}} - b~R_{\rm ISCO}\frac{C_{\rm NS}}{\eta} + c ,0\right) \right]^{1+d}.
\end{equation}

Here, $\eta$ = $Q/(1+Q^{2})$ is the symmetric mass ratio, $\hat{M}$ = $M^{\rm NSBH}_{\rm rem}/M^{\rm b}_{\rm NS}$, and $a=0.40642158$, $b=0.13885773$, $c=0.25512517$, $d=0.761250847$. The disk mass $M^{\rm NSBH}_{\rm disk}$ is calculated by subtracting $M^{\rm NSBH}_{\rm dyn}$ from $\hat{M}^{\rm NSBH}_{\rm rem}$. Of course, if the remnant mass is zero, this also corresponds to no tidal disruption. 
We draw $\zeta$ from a Gaussian centered at 0.3 with width $\sigma=0.15$, and multiply it by $M^{\rm NSBH}_{\rm disk}$ to obtain the wind ejecta mass $M^{\rm NSBH}_{\rm wind}$. Following \citealt{Barbieri_2019,2024MNRAS.52711053M} we further require $M_{\rm dyn}$ to be less than 50\% of $M_{\rm rem}$, to be consistent with results found by numerical simulations. For BNS (NSBH) mergers, an ejecta mass cut-off point of $10^{-5}~M_\odot$ ($10^{-4}~M_\odot$) was used to consider the KN as produced.

We use these outflow properties to simulate KNe lightcurves for all BNS and NSBH events in $u,g,r,i,z,y$ bands. For $g$ and $i$ bands, we check whether the simulated KN is detectable for those events for which the pointings covered by our observing plans matches the true sky location of the event. This provides an estimate of the KNe that can be detected in these 2 bands using DECam during the O4 and O5 runs. Here, detection means at least one observation in the simulated lightcurve is above the $5\sigma$ depth. We use AB magnitudes throughout our work.

\subsection{Simulated Observing Strategy}\label{obs strategy}

We consider the case of optimizing EM counterpart searches for DECam, however, most of our results can be applied to other instruments with similar filters. DECam is a wide-field high-performance CCD camera mounted on the 4m Blanco Telecope at Cerro-Tololo Inter-American Observatory in Chile. We take DECam to have a base sensitivity of 24.3 AB mag in $g$-band and 23.9 AB mag in $i$-band (at $5\sigma$ depth at 7 days from new Moon) for our fiducial strategy. To simulate follow-up observations, we use the Gravitational-Wave Electromagnetic Optimization code (\texttt{gwemopt}; \citealt{Coughlin_2016,Coughlin_2018gwemopt}) with DECam tiling. 

We calculate the exposure time (ET) in $g$ and $i$ bands (ET$_{\rm g}$ and ET$_{\rm i}$ respectively) as the time needed to reach a $5\sigma$ depth at the magnitude given by the 90th quantile of the KN magnitudes in our simulations. We use magnitudes at 1 day from the merger for BNS events and at 2 days from the merger for NSBH events since this is when the KN magnitude peaks. This is done in bins of luminosity distance. For the calculated exposure times, we then compute the number of tiles $N_{\rm t}$ that can be observed as:

 \begin{equation}
     N_{\rm t} = \frac{4~\text{hours}}{ET_{\rm g} + ET_{\rm i} + 60~\text{seconds}},
 \end{equation}
where we assume a maximum observing time of 4 hours and a telescope overhead time of 60s (30s per exposure). This simulated follow-up requires the airmass to be less than 2.2, and factors including the field of view and sensitivity of the telescope are taken into consideration. Our final goal is to use the simulations to choose an ideal observing time that is not necessarily similar to that of what we consider our fiducial strategy throughout this work, but that maximizes our chances of detecting a KN at a specific distance.

\section{Results}\label{section:Results}
\begin{table*}[!ht]
\begin{center}
\begin{tabular}{c | c | c | c | c | c | c | c | c | c}
\hline
\hline
Network & Binary type  & $A<1000$  & $A<500$  & $A<100$ & $A<10$  & $V_{90}<10^{3}$ & $A_{\rm 90_{\rm median}}$ & $V_{\rm 90_{\rm median}}$ & $D_{\rm median}$\\

& & deg$^{2}$ & deg$^{2}$ & deg$^{2}$ & deg$^{2}$ & ${\rm Mpc}^{3}$ & ${\rm deg}^{2}$ & $10^{6} {\rm Mpc}^{3}$ & ${\rm Mpc}$ \\
\hline
HL O4 & BNS & 22\% & 8.8\% & 1.0\% & 0\% & 0.10\% & 1820 & 19 & 259\\
& NSBH &  19\% & 7.7\% & 0.90\% & 0\% &  0\% & 2110 & 83 & 384\\
\hline
HLV O4 & BNS & 28 \% & 14\% & 3.8\% & 0.59\% & 0.47\% & 1670 & 22 & 276\\
& NSBH & 27\% & 15\% & 4.5\% & 0.46\% & 0.10\% & 1930 & 96 & 416\\
\hline 
HLVK O5 & BNS & 37\% & 26\% & 13\% & 2.9\% & 0.53\% & 1690 & 220 & 614\\
& NSBH & 39\% & 28\% &  14\% & 3.4\% & 0.34\% & 1740 & 610 & 901\\
\hline
\hline
\end{tabular}
\end{center}
\caption{Percentage of events with localization precision below a given threshold, for the different network configurations considered. Columns 3-6 show the percentage of detected GW  events with 90\% CI localization area below 1000, 500, 100, and 10 deg$^2$, respectively, column 7 shows the percentage of events with a 90\% CI volume below 10$^3$ Mpc$^3$, columns 8, 9 and 10 show the median values of the 90 CI area, volume and luminosity distance respectively.}
\label{tab:fractions}
\end{table*}

\begin{figure*}[!ht]
    \includegraphics[scale=0.55]{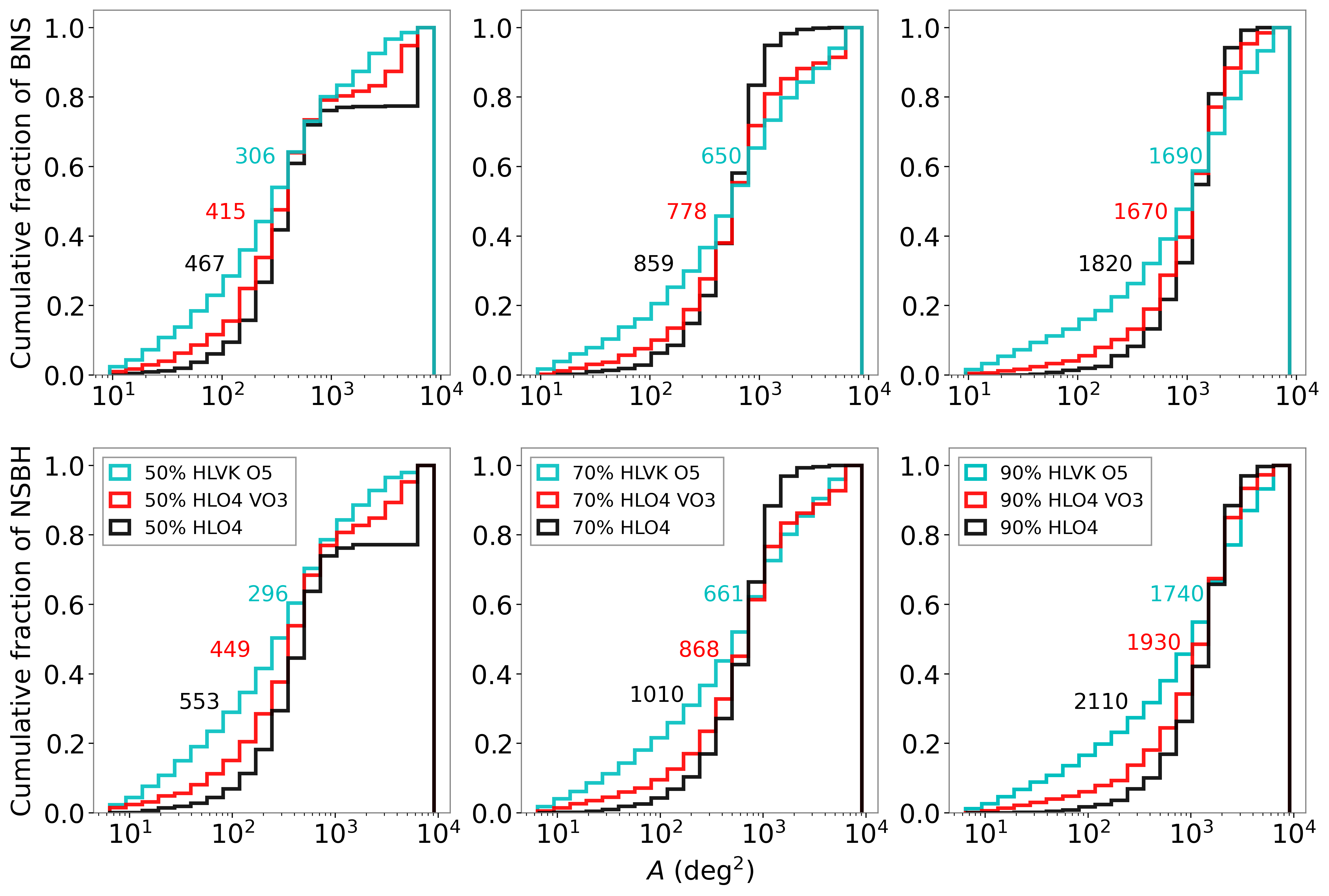}   
    \caption{Cumulative distribution of sky localization area at 50, 70, and 90$\%$ credible intervals for simulations 1 (black line), 2 (red line), and 3 (blue line) for BNS (top) and NSBH (bottom) events. The median localization area for each configuration is annotated in the respective panel.}
    \label{fig:w_Virgo}
\end{figure*}

\subsection{GW Localizations and rates}
\begin{figure}[!ht]
    \includegraphics[scale=0.345]{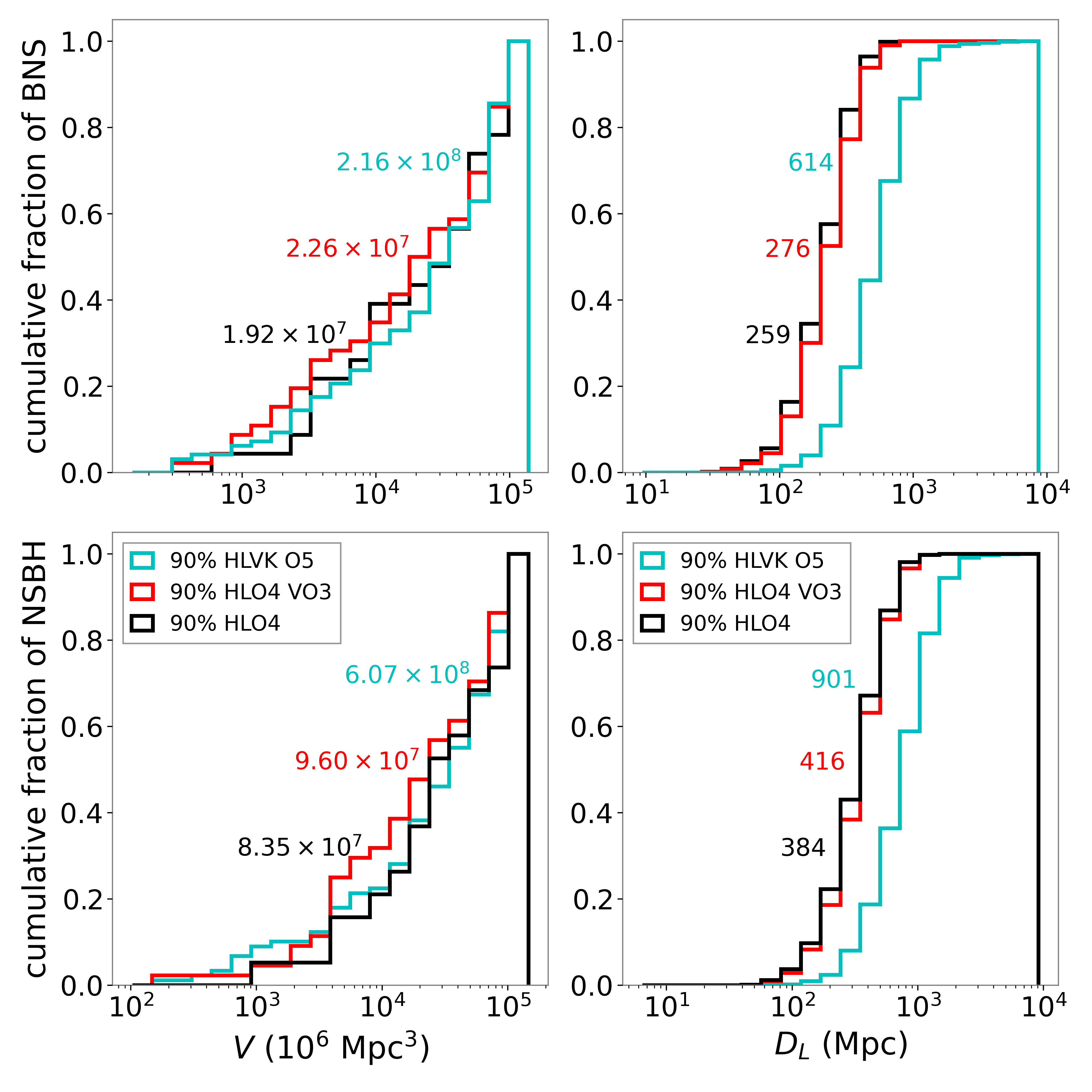}   
    \caption{Cumulative distribution of volume and distance for BNS (top) and NSBH (bottom) at 90$\%$ credible intervals for the 3 simulations using the same color scheme as in Figure \ref{fig:w_Virgo}. The median distance or volume for each configuration is annotated in the respective panel.}
    \label{fig:vol_dis}
\end{figure}

In this work, we focus on the results for BNS and NSBH mergers from our simulations, although binary black hole simulations are also produced. Since the GW source localization is determined by the sky area and the comoving volume contained within some probability, as well as the luminosity distance, we show cumulative distributions of these quantities for the different simulations considered in Figure \ref{fig:w_Virgo} and \ref{fig:vol_dis}. At the $90\%$ credible level, the median sky localization area for BNS events does not change significantly between O4 and O5. For NSBH systems, the median area slightly increases from $1,740~\mathrm{deg}^{2}$ in O4 to $1,930~\mathrm{deg}^{2}$ in O5 (Figure \ref{fig:w_Virgo}). As shown in Figure \ref{fig:vol_dis}, the median localization volume is of the order of $10^{7}$ Mpc$^{3}$ for O4 simulations and of the order of $10^{8}~\mathrm{Mpc}^3$ for O5, as a result of the increased distance reach. The median luminosity distance for BNS events increases from 300~Mpc in O4 to 600~Mpc in O5, while for NSBH systems, the median distance goes from 400~Mpc to 900~Mpc.

\begin{figure}
        \centering
\includegraphics[width=0.48\textwidth]{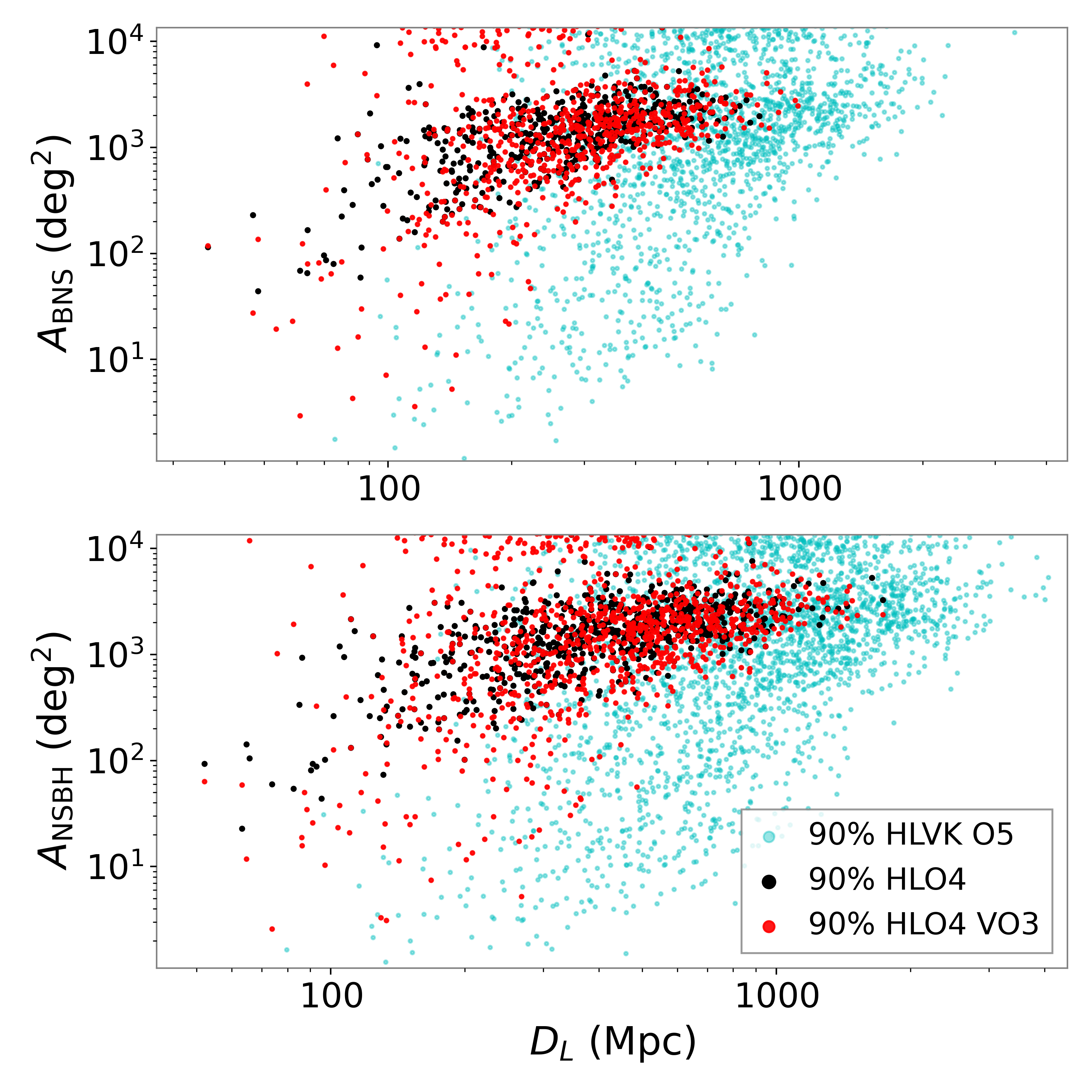}
        \caption{90$\%$ CI area versus luminosity distance for BNS (top) and NSBH (bottom) events with the same color scheme as in Figure~\ref{fig:w_Virgo}. }
        \label{fig:area_dist}
 \end{figure}

We also report the percentage of well-localized events (based on area and volume) in Table \ref{tab:fractions}. As expected, the inclusion of Virgo even at the sensitivity of O3 results in the detection of a higher number of well-localized events (90\% CI area $< 100$ deg$^{2}$) at distances $<100$ Mpc as can also be seen in Figure \ref{fig:area_dist} based on area, and in Figure \ref{fig:vol_dis} based on localization volume. For example, as shown in Table \ref{tab:fractions} events with 90\% CI area within 100 sq. deg. see a four-fold increase in numbers with the inclusion of Virgo.
This is in part because increasing the number of detectors would positively impact event localization due to the slight difference in arrival time of the signal. This is crucial for typical follow-up campaigns, which may focus on the better localized events by imposing a cut in area. Less distant events produce higher SNRs and, therefore, smaller localization regions compared to further events with the same chirp mass. Thus, as we go to larger areas, we probe more distant events, making the improvement in localization less prominent. For O4 with Virgo we find that events localized to within 100 sq. deg. occur at distances $< 220$ Mpc ($< 490$ Mpc), while those localized to $<500$ sq.\ deg.\ occur at distances $<360$ Mpc ( $<580$ Mpc) for BNS (NSBH).

Compared to the two O4 detector configurations, the O5 network detects events out to larger distances and detects a larger number of well-localized events, both in terms of area and volume. As expected, due to the larger distances of most detections, the median volume of events appear larger than or similar to O4, despite the growing number of detectors. In O5, events localized to within 100 sq. deg. occur at distances $<960$ Mpc ($<1400$ Mpc), while those localized to $<500$ sq.\ deg.\ occur at distances $<1100$ Mpc ($<2400$ Mpc) for BNS (NSBH).

We assume a volumetric merger rate of $170^{+270}_{-120}~\mathrm{yr}^{-1}~\mathrm{Gpc}^{-3}$ for BNSs,
and $27^{+31}_{-17}~\mathrm{yr}^{-1}~\mathrm{Gpc}^{-3}$ for NSBHs, following the merger rates in Table II of \citealt{abbott_population_2023}, where NS and BHs in this case are delineated at 2.5 $M_{\odot}$. The volumetric merger rate is more precise for BNS events compared to NSBH events likely due to the higher number of NSBH detections compared to BNS detections. We expect $\sim 2-20$ ($\sim 2-10$) BNS (NSBH) detections per year in O4, and $31-270$ ($19-110$) BNS (NSBH) detections per year in O5. These rates are obtained based on observations during the first three observing runs. We update these predictions based on the fact that no high-significance BNS alert has been issued in O4 up to the end of O4b. 

Following this, the updated 90\% CI rate interval for BNSs in the PDB model becomes $14-124~\mathrm{yr}^{-1}~\mathrm{Gpc}^{-3}$ based on O4b. Therefore, our preliminary GW BNS detection rates based on the non-detection upto O4b is $ 1-6 ~\mathrm{yr}^{-1}$ for O4 and $9-76~\mathrm{yr}^{-1}$ for O5.

In Table \ref{tab:fractions} we also report the fraction of events with a volume localization of $<1000$ Mpc$^3$, corresponding to a volume within which we expect to find less than one galaxy cluster ($\sim 0.1$ based on \citealt{Redmapper}) with halo mass $\gtrsim 10^{13.5}~M_\odot$. This is interesting because short gamma-ray bursts have also been observed to occur in galaxy clusters (e.g., \citealt{Nugent_2020}), and GW170817 occurred in a galaxy group. Although such well-localized events are extremely rare, it is possible that one of these could be observed in O5 given the yearly detection rates combined with the potentially long (few-year) duration of O5.

\subsection{Kilonova Detectability}

\begin{figure*}[htpb!]
    \includegraphics[width=0.53\linewidth]{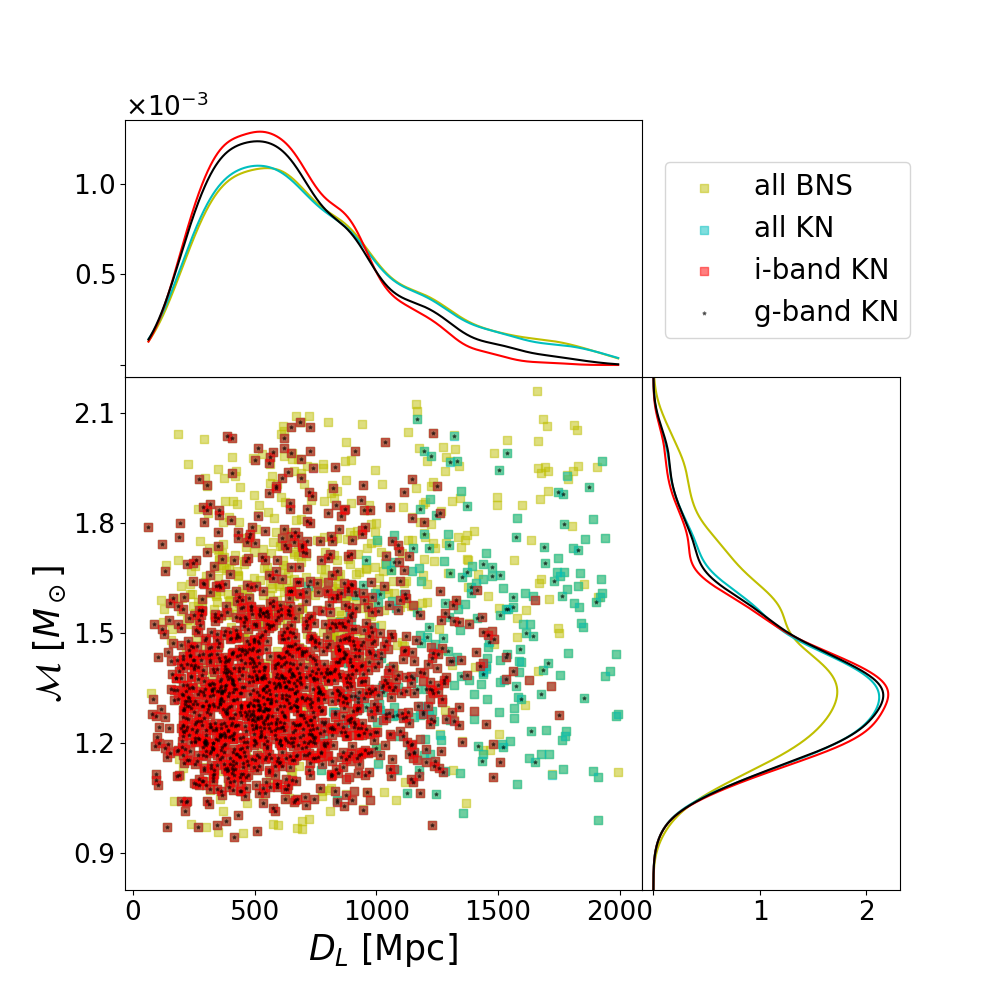}\includegraphics[width=0.485\linewidth]{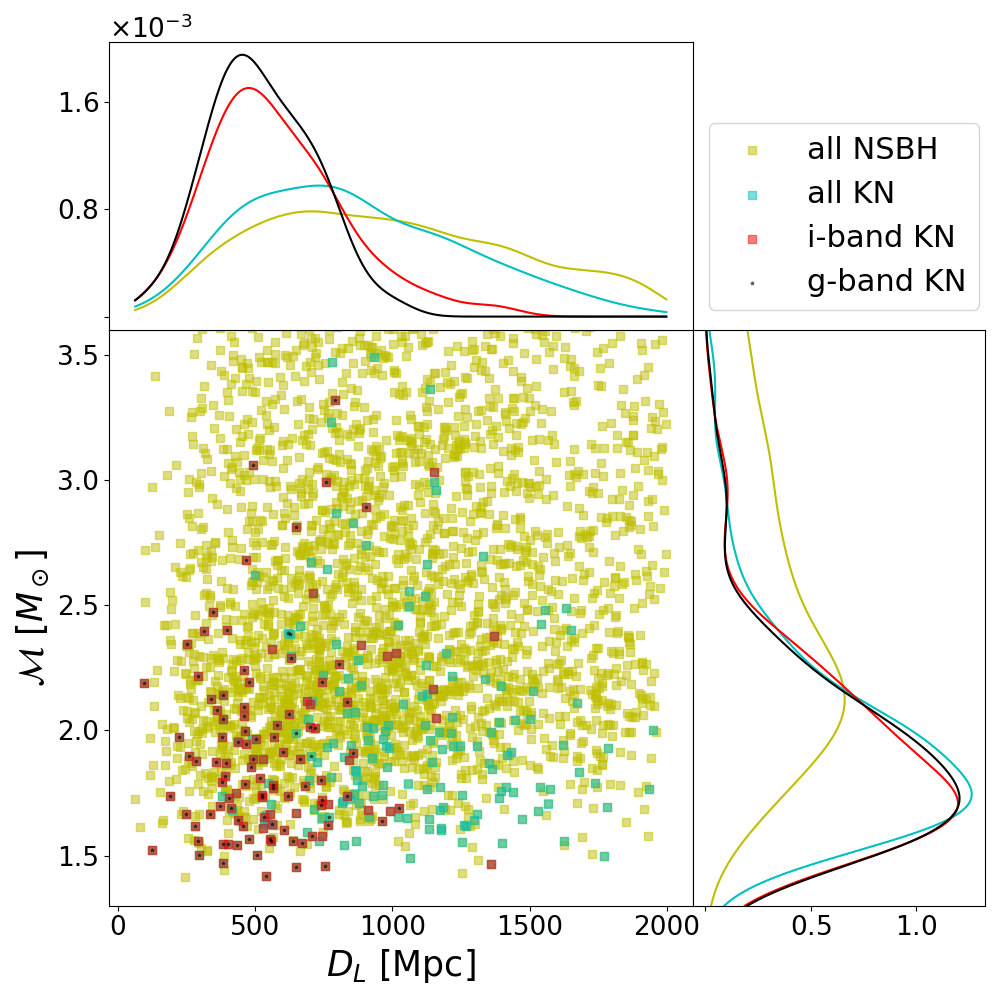}
    \caption{Chirp mass versus luminosity distance for BNS (left) and NSBH (right) events in the O5 simulation. The yellow squares represent all BNS/NSBH detected in gravitational waves, the cyan squares show all mergers for which a KN is produced according to our model and fiducial EOS. Out of all the KNe, those with a black point and/or a red circle were detected in $g$ and/or $i$ band, respectively. The same color scheme applies to the histograms. We limit the plot to a chirp mass of $3.5~{M_{\odot}}$ for the NSBH since no KN is produced beyond this chirp mass in our simulations. For both cases, the lower chirp mass end of the distribution yields higher KN detection rates than at larger mass.}
    \label{fig:tm_dis}
\end{figure*}

\begin{figure*}[htpb!]
    \includegraphics[width=0.53\linewidth]{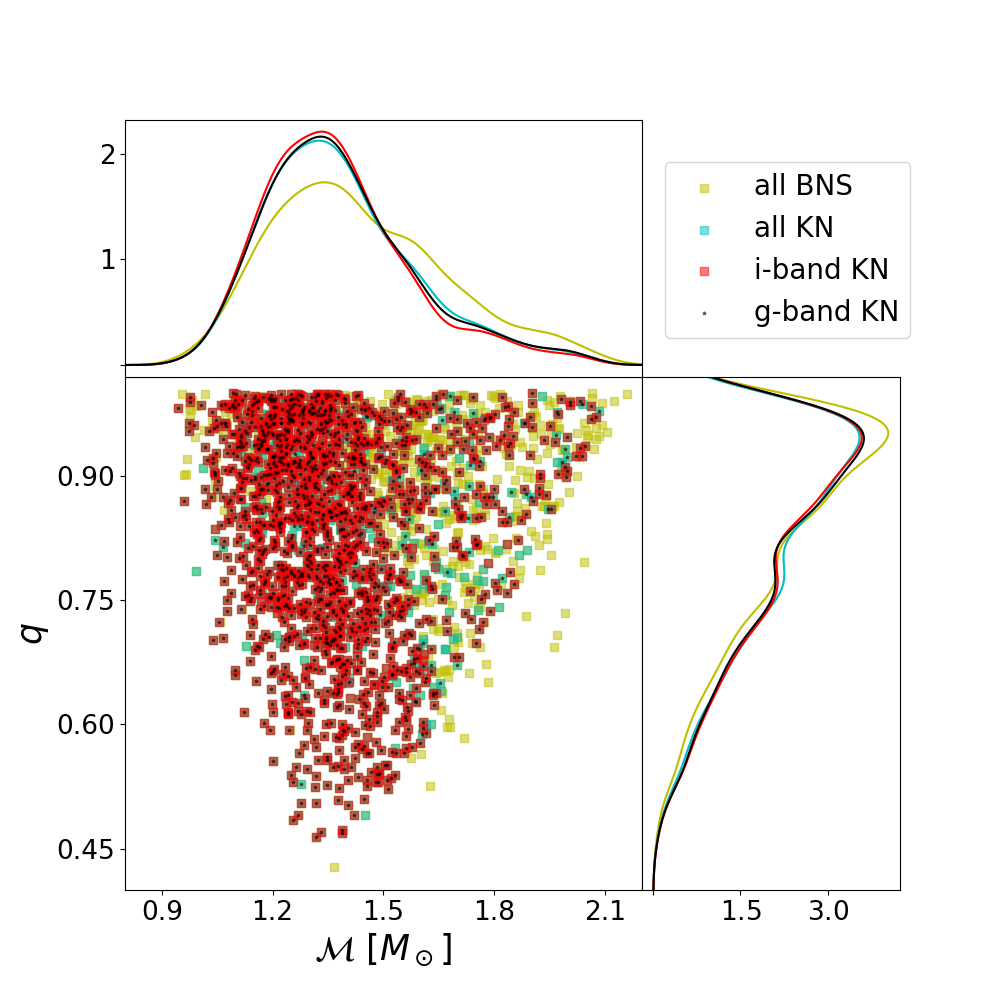}\includegraphics[width=0.485\linewidth]{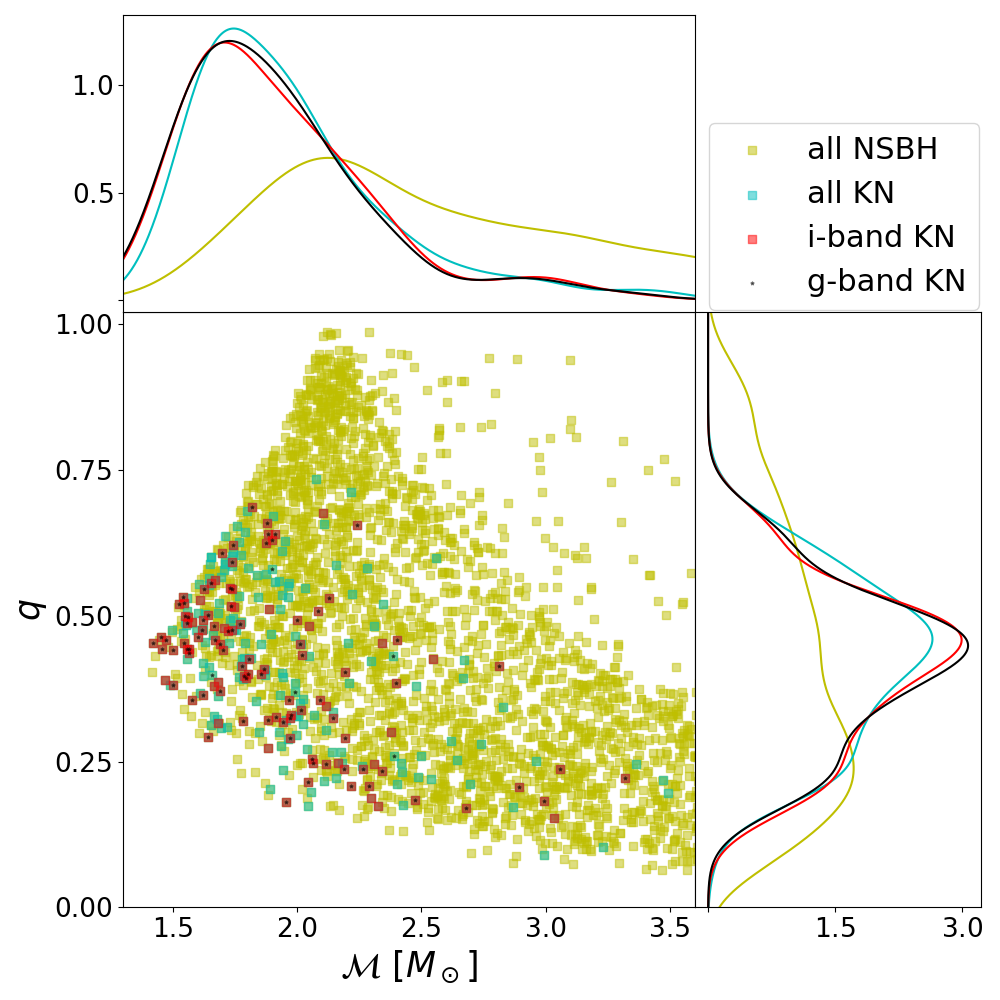}
\caption{Mass ratio versus chirp mass for the BNS merger events (left panel) and NSBH mergers (right panel) from the O5 simulation. The color scheme is the same as in Figure \ref{fig:tm_dis}.}
\label{fig:mratio_tm}
    \end{figure*}

In what follows, we explore the KN detectability dependence on various binary observables in order to inform our follow-up strategy. At first, we do not consider coverage by \texttt{gwemopt} (which is considered in the next subsection); but rather we consider the event as  detectable if the KN lightcurve in a given filter is brighter than the DECam detection limit (24.3 mag in $g$-band and 23.9 mag in $i$-band for the fiducial exposure time). We choose to show the results from the O5 simulations in the figures as these include a larger number of objects, and the corresponding O4 plots would be mostly repetitive. Note that in what follows, O4 is used to denote simulation~2 and O5 is used to denote simulation~3. Two quantities that can typically be easily constrained from the GW data are chirp mass and distance, so we start by focusing on combinations of these parameters. 

\subsubsection{Chirp mass dependence}
In Figure \ref{fig:tm_dis} we show the chirp mass versus luminosity distance for all events that we classify as BNS (left) or NSBH (right) in our O5 simulations. We limit the luminosity distance to less than 2000 Mpc since the KN detectable with the projected O4/O5 sensitivity is well within this distance range. Because our ejecta masses, $M_{\rm dyn}$ and $M_{\rm wind}$, are allowed to reach zero, not all BNSs and NSBHs detected in GWs will have an associated KN. We find that for the fiducial EOS, $\sim 78\% (77\%)$ of our BNS events and $\sim 6\% (7\%)$ of our NSBH events have an associated KN with either $M_{\rm dyn}$ or $M_{\rm wind}$ $>$ 10$^{-5}$ M$_{\odot}$ for BNS and $M_{\rm dyn}$ or $M_{\rm wind}$ $>$ 10$^{-4}$ M$_{\odot}$ for NSBH in O4 (O5). \\

One of the many factors that favor KN detection is a lower chirp mass (hence lower total mass) of the merging binaries, which is expected, at least for the chosen EOS, as lower-mass stars are typically more easily disrupted. An interesting point to note is that the chirp mass appears to be a more significant discriminator between detection/non-detection than the luminosity distance, at least within the O4 and O5 distances probed by our simulations. This is because even at distances $< 1$ Gpc, a high chirp mass may result in no kilonova production. This is especially evident in the NSBH scenario where higher chirp mass would imply that the NS is engulfed before significant tidal disruption. Moreover, due to its higher sensitivity, $g$-band lets us detect BNS KNe at $>1000$ Mpc, slightly better than $i$-band, as shown in the top distribution of Figure \ref{fig:tm_dis}. The 90th percentile of KN detected from BNS merger are expected to have a chirp mass of $\lesssim 1.7~M_\odot$ ($\lesssim 1.6~M_\odot$) and are expected to be at a distance $\lesssim 1140$ Mpc ( $\lesssim 1040$ Mpc) in $g$-band ($i$-band).

On the other hand, NSBH KN detections peak at $\sim 460$ Mpc in $g$ band, but for distances $\gtrsim$ 780 Mpc, NSBH KNe are more often detected in $i$ band than in $g$ band, which is reasonable considering the larger redshifts and the fact that NSBHs can be red due to a small squeezed $M_{\rm dyn}$ component and the red wind ejecta component \citep{Kasen_2017}. In fact, the different dynamical ejecta geometry and composition is taken into account in the KN models used: the NSBH KN models do not feature the lanthanide-free polar component assumed for the BNS KNe, thus producing intrinsically redder transients than the BNS KNe \citep{Dietrich_2020,Anand_2020}. The 90th percentile of the detectable KNe from O5 NSBH mergers is expected to have chirp mass $\lesssim 2.4~M_\odot$ and to be at a distance $\lesssim 765$ Mpc ($\lesssim 905$ Mpc) in $g$-band ($i$-band).
\begin{figure*}[htpb!]
    \includegraphics[width=0.53\linewidth]{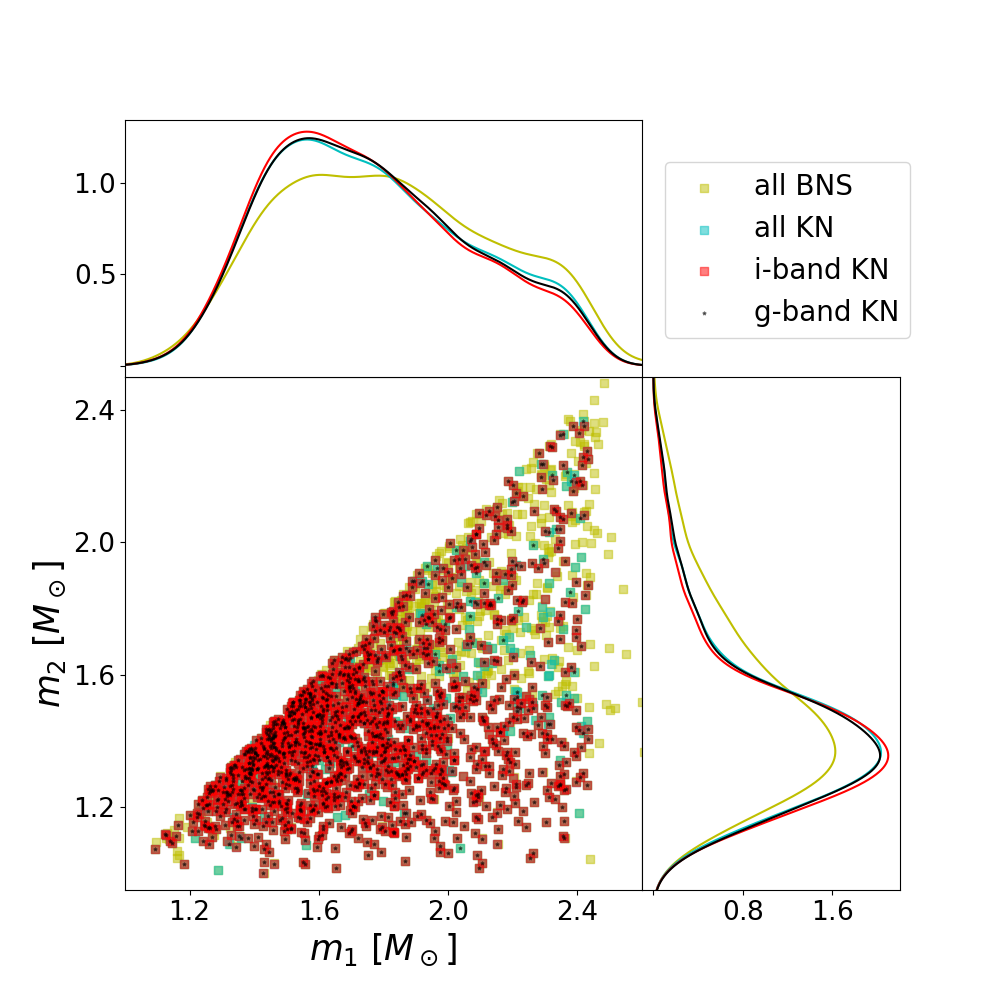}
    \includegraphics[width=0.485\linewidth]{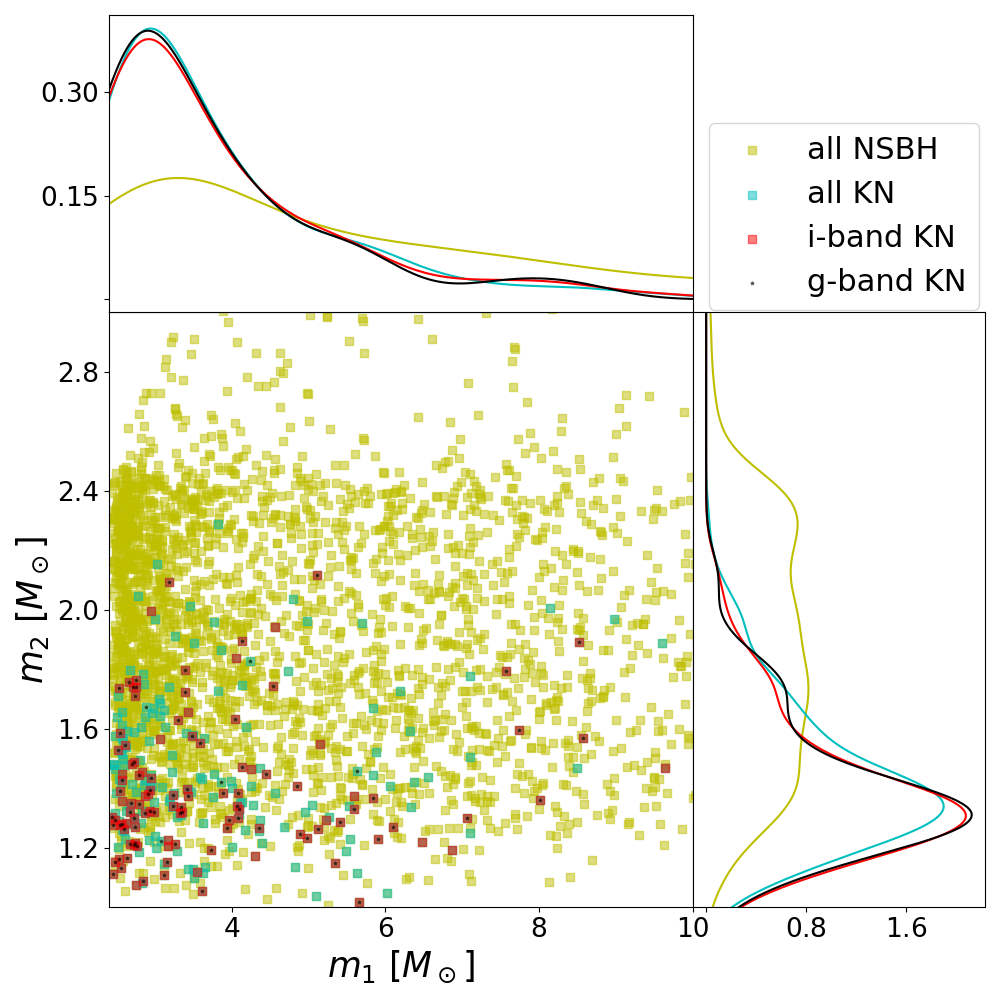}
\caption{Component masses of BNS (left) and  NSBH (right) systems and their effect on KN detection. We limit $m_1$ to $8_{M_{\odot}}$ in the plot for visualization purposes. The color scheme is the same as in Figure \ref{fig:tm_dis}.}
\label{fig:m1m2}
    \end{figure*}

\subsubsection{Mass ratio dependence}

In Figure \ref{fig:mratio_tm}, we show the mass ratio and chirp mass of the BNS and NSBH systems for which GWs are detectable, and the subset for which a KN is also detectable. Although the chirp mass is a good proxy for KN detection efficiency, the mass ratio also provides a good discriminator. Although gravitational wave detectors are more sensitive to symmetric-mass binaries, kilonova detection favors asymmetric systems due to their greater likelihood of producing significant ejecta. Among GW detected BNS mergers, $\sim 77\%$ of systems with $q < 0.8$ have detectable kilonovae, compared to 65\% for symmetric systems ($q > 0.8$). While kilonovae from symmetric binaries are similarly likely to be detectable once produced ($86-90\%$ of kilonovae are detectable for mass ratios greater or smaller than 0.8), they are less likely to generate kilonova emission in the first place given our ejecta mass requirements.

\begin{figure*}
\centering
\includegraphics[width=0.53\linewidth]{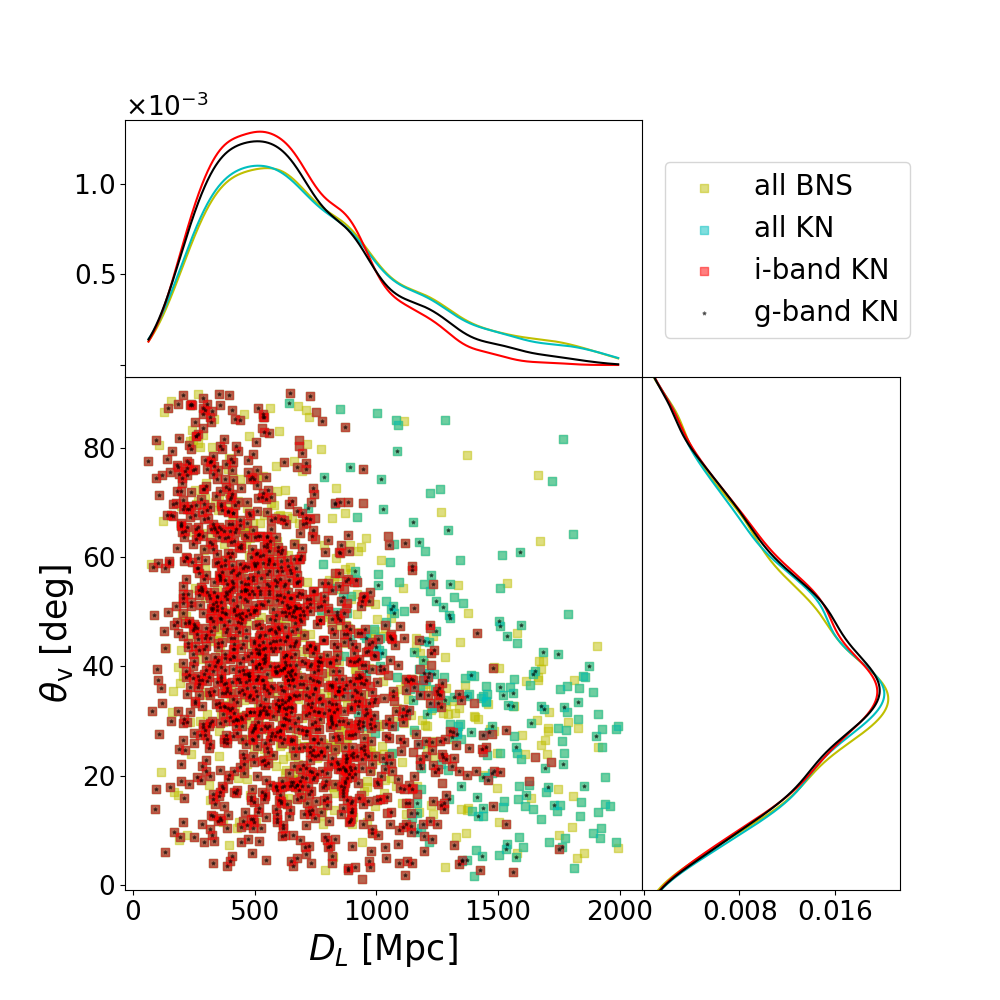}\includegraphics[width=0.485\linewidth]{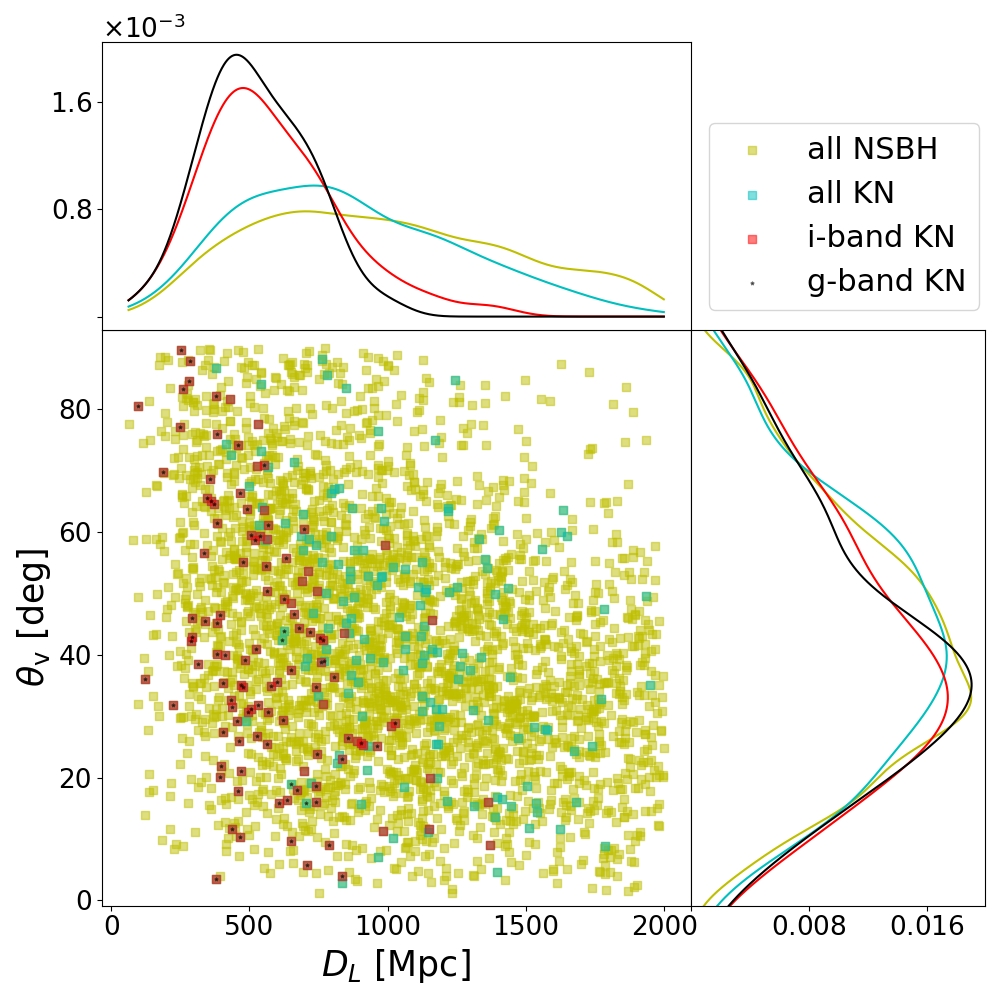}
\caption{Viewing angle of the BNS (left) and the NSBH (right) mergers versus their distance for the O5 simulations. The color scheme is the same as in Figure \ref{fig:tm_dis}.}
\label{fig:view_dis_bns}
\end{figure*}
For NSBH mergers, the mass ratios follow the  distribution shown in Figure~\ref{fig:mratio_tm}. We interpret the peak at $q \sim$ 0.5 with NSBHs consisting of a BH close to the low-mass edge of the dip in the PDB model (where the kilonova production and detectability peaks as expected). For masses below that edge, there are close to no BHs (although note the transition is smooth in our model), while in the dip, the merger rate decays, hence the peak in the observed mass ratio distribution. Because KN production is favored for low-mass BHs and low-mass NSs, they can only be detected at lower distances. For these reasons, the 90th percentile of detectable NSBH KNe lies in a specific region of the mass ratio-distance space ($q \in[0.2-0.7]$ and distance $<1$ Gpc). The rest of the NSBH mass ratio distribution follows from the extension of the BH mass distribution out to masses much larger than the NS mass. KN production is highly suppressed in this regime ($q < 0.1$) since the highly asymmetric mass ratio would imply that the BH is large enough to engulf the NS before it is tidally disrupted.

\subsubsection{Component mass dependence}

Compared to the distribution of GW BNS events, those that result in the detection of a KN clearly shows a preference towards lower values of $m_2$ as shown in Figure \ref{fig:m1m2}. For BNS mergers, the KN detection efficiency peaks around $m_1\sim 1.6 M_{\odot}$, with events having $m_1 \leq$ 1.6 $M_{\odot}$ detected with an efficiency of 89\% in both $g$ and $i$ bands. For $m_2$, the detection efficiency peaks at 1.3 $M_{\odot}$, where events with $m_2 \leq$ 1.3 $M_{\odot}$ are detectable at an efficiency of 91\% (92\%) in $g$ ($i$) band respectively. As a result of the increased compactness of the secondary component at higher masses, tidal disruption would not be strong enough to produce significant ejecta mass. The compactness and mass are indeed parameters that explicitly enter in our $M_{\rm dyn}$ Eq. \ref{eq:mdyn_bns} prescription, so the existence of a KN detectability dependence on secondary mass is not surprising. The higher sensitivity in $g$-band compared to the $i$-band makes it easier to detect some of these events in the former, simply because these events tend to be at larger distances than those closer to the peak of the mass distribution (higher mass BNSs have a lower volumetric merger rate than lower mass NSs in our model). 

In the NSBH case, the compactness of the NS also determines the launch of ejecta (again, it is explicitly entered in our $M_{\rm dyn}$ Eq. \ref{eq:M_dyn_nsbh} and $M_{\rm wind}$ Eq. \ref{eq:mdisk_nsbh} prescriptions). Of course, the BH mass also plays an important role in defining the ISCO radius. In our simulations, the 90th percentile black hole mass for all NSBH mergers that produce a kilonova—as well as those with detectable KNe emission in both $g$ and $i$ bands are below $6 M_\odot$. Even though the astrophysical BH distribution we assume \citet{Farah_2022} has a dip around $2.2 - 6 M_{\odot}$, we do not see a noticeable decrease in number of detected NSBH with $m_1$ around $2.2-6~M_\odot$ compared to a larger number of NSBHs with primary mass above this range. This is because the pairing function assumed in \citet{Farah_2022} to construct the NSBH population favors nearly equal mass binaries. As a result, in our simulation, BHs with mass above the dip are less likely to be in NSBH systems, and they mostly form binary black hole systems.

\begin{figure}[htpb!]
    \centering
    \includegraphics[width=\linewidth]{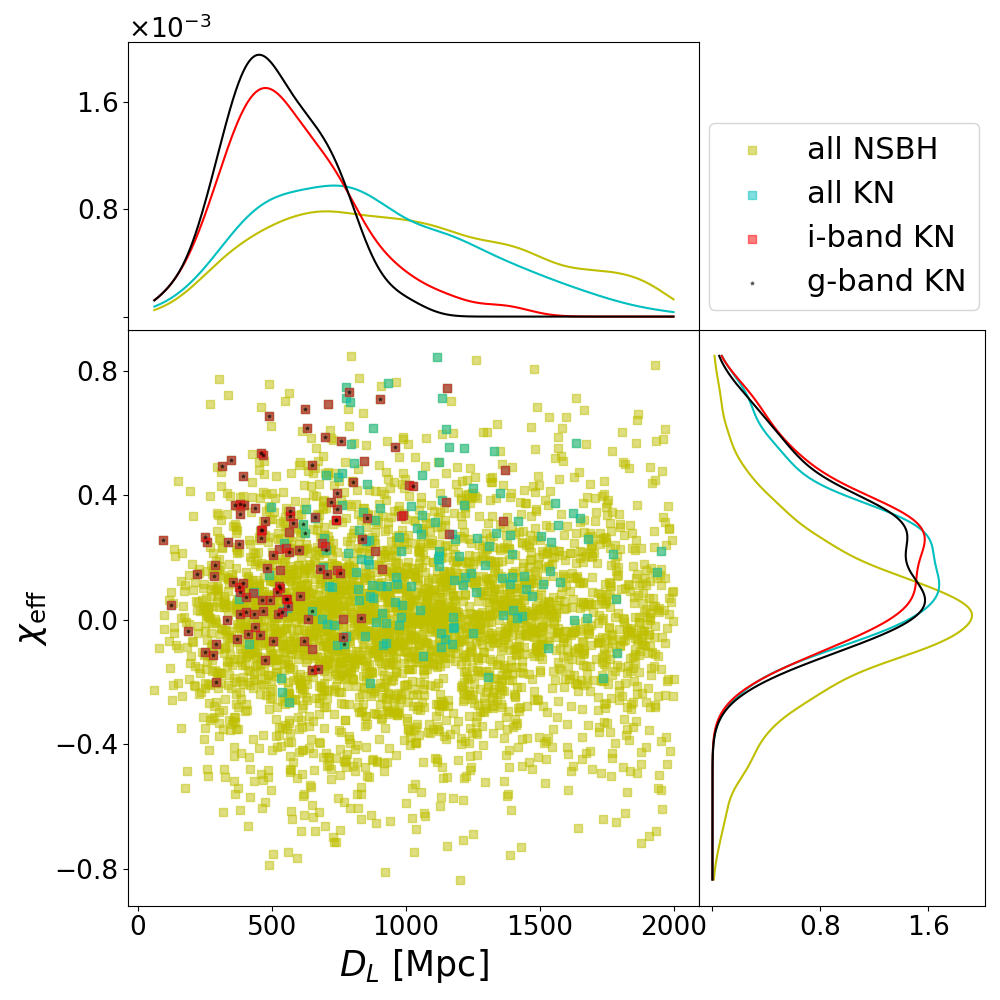}
    \caption{Effective spin $\chi_{\rm eff}$  versus luminosity distance for our O5 simulated NSBH mergers. The color scheme is the same as in Figure \ref{fig:tm_dis}.}
    \label{fig:effspin_nsbh}
\end{figure}

\begin{figure*}[htpb!]
    \centering
    \includegraphics[scale=0.3]{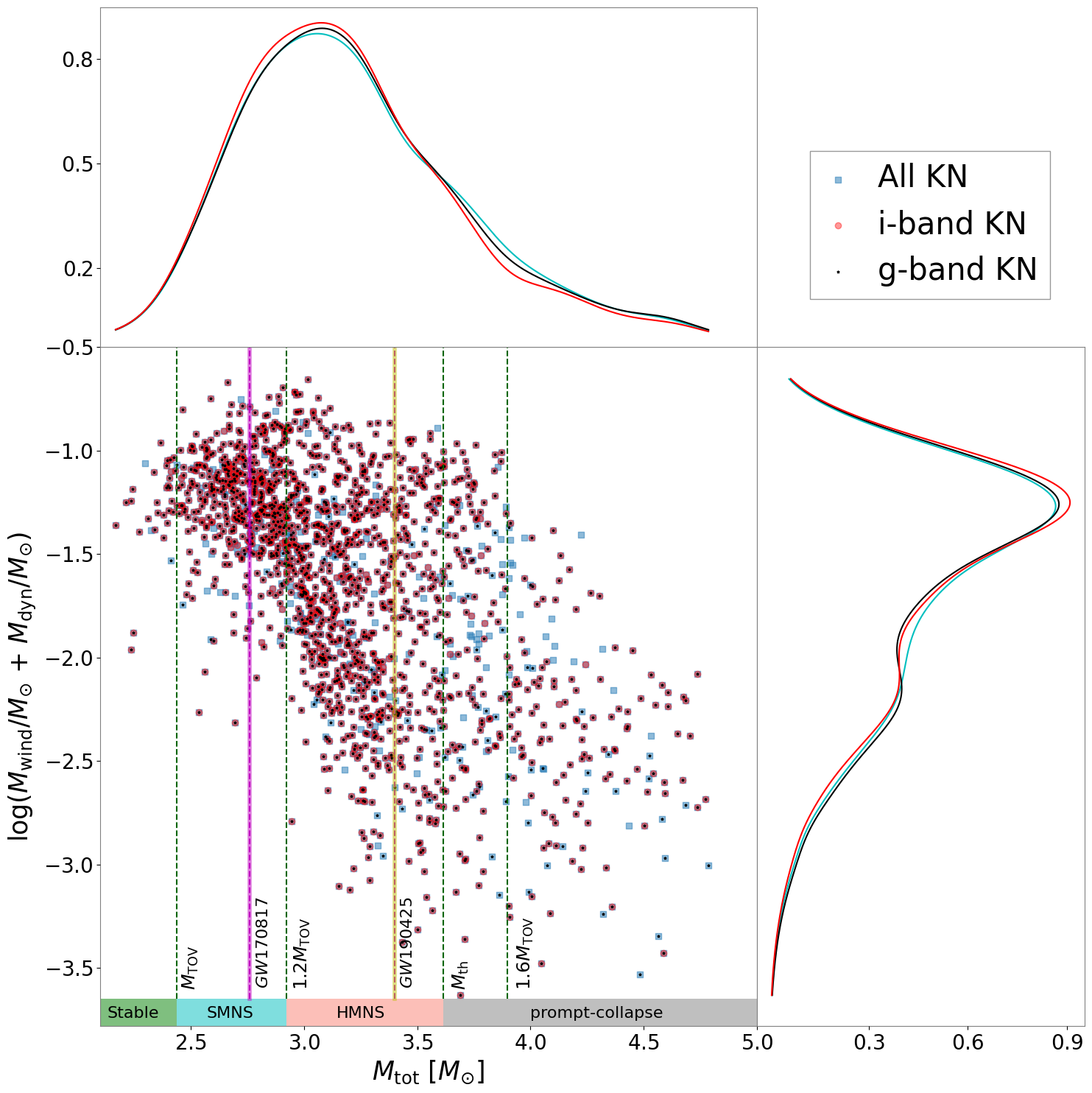}
    \caption{Total ejected mass from BNS mergers versus their total mass. Vertical dashed lines delineate between different merger remnants fates, with the threshold masses and radii derived assuming the EOS from \citet{Huth:2021bsp}. Here $M_{\rm TOV}=2.436$ $M_{\odot}$, $R_{\rm TOV}=11.7$ km, and $M_{\rm th}=3.616$ $M_{\odot}$. The color scheme is the same as in Figure \ref{fig:tm_dis}, except that by construction we do not have BNSs that do not produce a KN here. The pink and yellow lines indicate the fate of GW170817 and GW190425 remnants according to our assumptions.
    }
    \label{fig:remnant fate}
\end{figure*}

\begin{table*}[!htpb]
    \setlength{\tabcolsep}{3pt}
    \centering
    \begin{tabular}{c|c|cc|cc|cc}
        \hline
        \multicolumn{1}{c|}{Remnant} & \multicolumn{1}{c|}{Binary mass range} & \multicolumn{2}{c|}{GW BNS detections} & \multicolumn{4}{c}{Detectable KNe} \\
        \hline
        & & O4 & O5 & \multicolumn{2}{c}{O4}  & \multicolumn{2}{c}{O5} \\
        \hline
        & & & & $g$ & $i$ & $g$ & $i$ \\
        \hline\hline
        Stable NS & $M_{\rm tot} < M_{\rm TOV}$ & 2.0\% & 1.9\% & 2.0\% & 2.0\% & 1.9\% & 2.0\% \\ 
        \hline
        SMNS & $M_{\rm TOV} < M_{\rm tot} < 1.2 M_{\rm TOV}$ & 27\% & 28\% & 27\% & 28\% & 28\% & 30\% \\
        \hline
        HMNS & $1.2 M_{\rm TOV} < M_{\rm tot} < M_{\rm th}$ & 49\% & 50\% & 49\% & 50\% & 51\% & 52\% \\
        \hline
        Prompt BH & $M_{\rm tot} > M_{\rm th}$ & 21\% & 20\% & 21\% & 20\% & 19\% & 17\% \\
        \hline\hline
    \end{tabular}
    \caption{Percentage of BNS mergers that falls into each of the remnant categories as predicted by the assumed fiducial EOS in LVK O4 and O5 runs. The ``BNS GW detections'' column refers to the percentage of GW BNS injections that are ``detected'' in GWs in our simulation and fall into each category. The ``detectable KNe'' column shows the fraction of objects with a detectable associated KN for each category, out of all objects with an associated KN.}
    \label{table bns remnant}
\end{table*} 

\subsubsection{Viewing angle dependence}

In Figure \ref{fig:view_dis_bns} we show the viewing angle of BNS binaries versus their distance. There is a marginal dependence of KN detectability on viewing angle: systems that are closer to on-axis are seen out to further distances than those more edge-on. This trend is more prominent in $g$-band than in $i$-band, particularly for face-on systems at larger distances. This is expected for BNS mergers, as the KN ejecta includes a lanthanide-poor polar component (driven by shock and neutrino-driven winds), which produces bluer and brighter emission along the polar axis \citep{10.1093/mnras/stv721, Bulla_2019, Dietrich_2020}. This component dominates the observed signal when the binary is viewed close to face-on.

For NSBH systems, as shown in Figure~\ref{fig:view_dis_bns}, we find that detectability is primarily driven by the binary's distance. The NSBH kilonova model we adopt (based on \citealt{Bulla_2019}) includes a planar, lanthanide-rich dynamical component, with no squeezed polar dynamical ejecta \citep{Kawaguchi_2016, Bulla_2019, Foucart_2023}, resulting in emission that is generally redder than in the BNS case. This leads to different viewing angle dependencies in the blue band: BNS kilonovae are more easily detected when viewed face-on, whereas NSBH kilonovae may appear brighter at intermediate viewing angles due to the geometry of their ejecta. Consequently, detectability in the $i$-band tends to exceed that in the $g$-band for NSBH mergers. For NSBH kilonovae within 1~Gpc, the detection efficiency varies modestly with viewing angle: face-on systems ($0^\circ$–$30^\circ$) yield detection efficiencies of 31\% in $g$ and 33\% in $i$, intermediate angles ($30^\circ$–$60^\circ$) increase to 46\% and 48\%, respectively, while edge-on systems ($60^\circ$–$90^\circ$) drop to 20\% in $g$ and 21\% in $i$.

\subsubsection{Spin dependence}
The spin of the BNS events does not have an impact on KN production and detection as it is not taken into account in our ejecta prescriptions, which is generally justified by the expected small dimensionless spin of the individual stars within BNS systems \citep{PhysRevD.98.043002}. Meanwhile, for NSBH mergers, the BH spin has a significant effect on KN production and detection. The radius of the ISCO of a BH is given by:
\begin{equation}
    R_{\rm ISCO} = f(\chi_{\rm BH})\frac{GM_{\rm BH}}{c^{2}},
    \label{risco}
\end{equation}
where $6\leq f(\chi_{\rm BH}) \leq 9$ for an orbit in which $\chi_{\rm BH}$ is anti-aligned with the direction of angular momentum (counter-rotating) and $1\leq f(\chi_{\rm BH}) \leq 6$ when $\chi_{\rm BH}$ is aligned with the direction of angular momentum (co-rotating) \citep{1972ApJ...178..347B,Sarin_2022}. For tidal disruption to occur, the disruption radius should be $r_{\rm disrupt} \geq R_{\rm ISCO}$. This means that tidal disruption is favored for systems with co-rotating BH spins. In Figure \ref{fig:effspin_nsbh} we show the dependence of a GW-observable, the effective spin $\chi_{\rm eff}=(m_{1}\chi_{1z} + m_{2}\chi_{2z})/(m_1+m_2)$ (where $\chi_{1z}, \chi_{2z}$ are the the components' spins along the binary angular momentum direction), on the detectability of events in our simulations, since the BH spin magnitude, which directly affects the ISCO radius, is typically harder to constrain from GW observations. Because the effective spin is mass-weighted, the BH component has a larger contribution to its value than the NS. We can see that both KN production and detection favor $ -0.2 < \chi_{\rm eff} < 0.7 $. This is a KN selection effect, not a GW one, as GW selection effects only slightly skew the effective spin distribution \citep{Ng_2018}.  

\subsubsection{Fate of the central remnant}
The fate of the central remnant and the properties of the ejecta are all related to the masses and spins of the components of the BNS system \citep{Shibata_2006}. In Figure \ref{fig:remnant fate} we show the total (wind and dynamical) ejecta mass in the simulated BNS mergers versus the summed (total) component masses, along with the expected mass ranges for the formation of different kinds of remnant. A hypermassive NS (HMNS) is a massive NS remnant, supported mainly by its differential rotation, that will collapse to a BH within tens to hundreds of milliseconds following the merger as mass accretion continues \citep{Siegel_2013}. A supramassive NS (SMNS) is a less massive NS supported by its solid body rotation and can survive a few hundreds of milliseconds until the remnant spin downs to a point of collapse \citep{metzger_kilonovae}. Above the threshold mass, we expect prompt collapse to a black hole. Note that GW170817-like events result in a SMNS remnant using our fiducial EOS, whereas with our softer EOS, such events are expected to produce a HMNS remnant. Previous work on GW170817 \citep{Gill_2019} excludes the possibility of a long-lived stable NS remnant, since a BH is most likely required \citep{Rezzolla_2015} to facilitate a relativistic outflow that would power a short GRB, but they could not exclude the possibility of a SMNS remnant with a magnetar-like surface dipole magnetic field powering a GRB \citep{Zhang_2001,Wei-Hong,10.1111/j.1365-2966.2008.12923.x} before it would ultimately collapse into a BH. Other works  \citep{10.1093/mnras/sty3047} mention the possibility of a remnant SMNS if $M_{\rm TOV}$ $\geq$ 2.16 $M_{\odot}$, which corresponds to the case of the fiducial EOS used in this work ($M_{\rm TOV}=$2.436 $M_{\odot}$). The possibility of a SMNS remnant is also explored by \citealt{Margalit_2017} whose work points to GW170817 producing either a HMNS or a short-lived SMNS. 

The threshold mass $M_{\rm th}$ depends primarily on the EOS used \citep{Hotokezaka_2011,Bauswein_2013}, but also on mass ratio. Symmetric mass ratio systems have higher $M_{\rm th}$ (by $\sim 0.1-0.3~M_\odot$ for a mass ratio of 0.6) compared to asymmetric-mass BNS systems. This follows from the fact that symmetric-mass binary systems have higher angular momentum at a given orbital separation compared to more asymmetric binaries assuming the same $M_{\rm tot}$ \citep{Bauswein_2017}, which facilitates stabilization of the remnant against collapse. For our fiducial EOS, $M_{\rm th}$ takes a value of 3.62 $M_{\odot}$, and we ignore the effect of mass ratio since our pairing function strongly disfavors asymmetric BNSs with $q<0.5$. 

We compare the proportion of each remnant category in the injected BNS population to the GW-EM detected sample. While the injected population is dominated by HMNS ($\sim 45\%$) and has a slightly lower fraction of prompt collapse systems (25\%--28\%), the GW-EM detected sample shows an even higher prevalence of HMNS (49\%--52\%) and a reduced prompt collapse fraction (17\%--21\%). This reduction arises because prompt collapse systems tend to have higher total masses, thus a lower probability of producing an observable kilonova. In contrast, HMNS, SMNS, and stable NS remnants, despite being less common in the injected population, have a significantly higher likelihood of producing bright electromagnetic counterparts, which boosts their representation in the KN-detected sample. As shown in Table \ref{table bns remnant}, most BNS we will detect in GWs are likely to result in an HMNS, but a noticeable fraction of multimessenger sources will promptly collapse into a BH. In our simulation, $\sim 2\%$ of both the BNS KN detections and the GW detections arise from a system that likely results in stable NS remnants, while we expect at least 27\% of the O4 KN detections to be from sources that would likely produce SMNS and at least 49\% from sources that result in HMNS remnants. This number comes to about 28\% in O5 for SMNS and 51\% for HMNS remnants. Around 20-21$\%$ of the KN detections in the O4 simulations and 17-19\% of KN detections from the O5 simulations are from sources that would undergo a prompt collapse. Although the finding that most BNS mergers will produce an HMNS is in agreement with \citealt{Margalit_2019}, they find that a negligible fraction of objects will undergo prompt collapse. This is a result of the different mass distributions assumed, as \citealt{Margalit_2019} uses a narrow symmetric mass distribution motivated by Galactic binaries. Such distribution is significantly lighter than the one used in this work, which accommodates for events such as GW190425 \citep{Abbott_2020_GW190425}, not easily explained by the Galactic binary population. 
\begin{figure}[t!]
    \centering
    \includegraphics[scale=0.66]{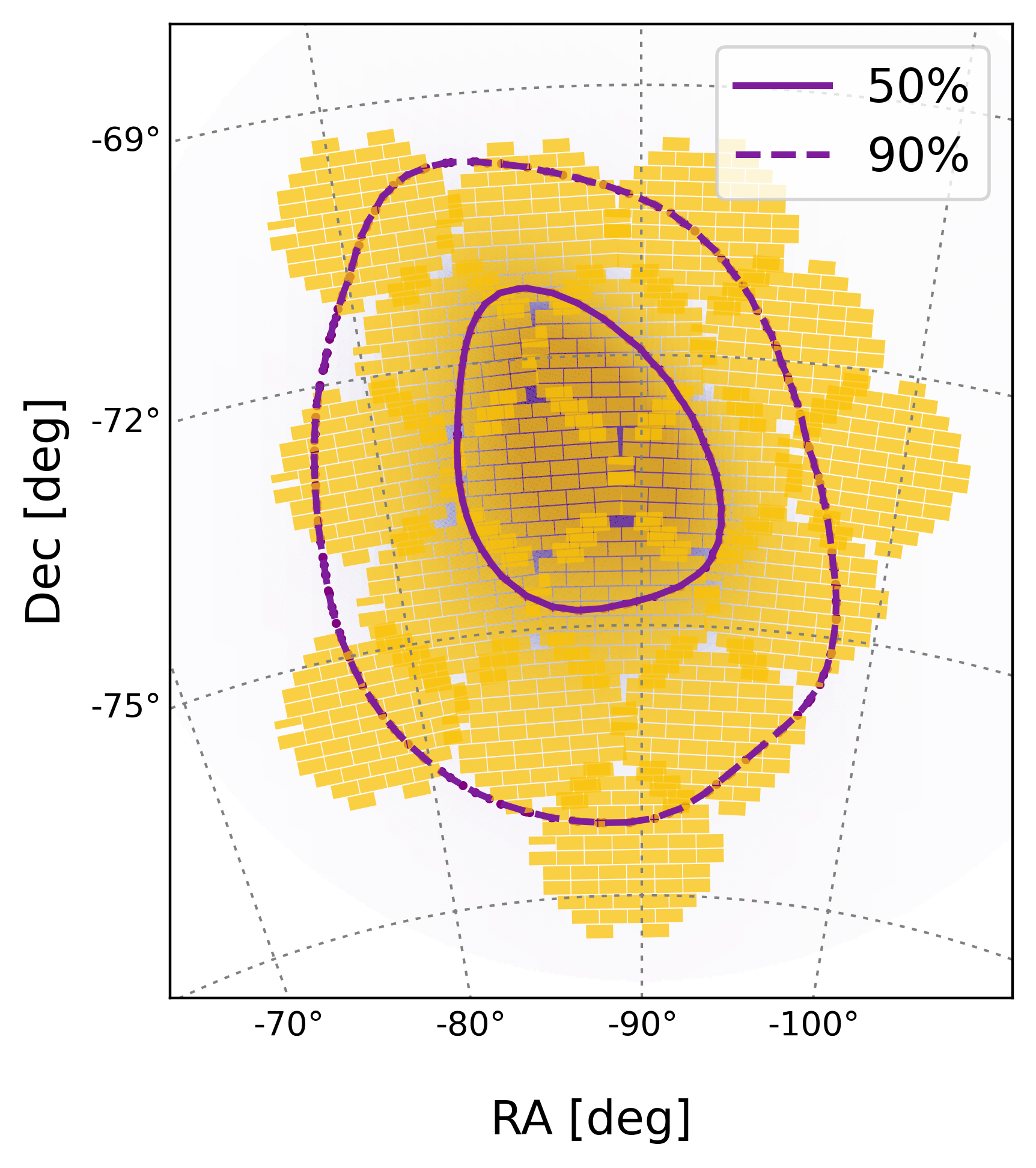}
    \caption{Simulated DECam tiling follow-up of a well-localized BNS event. The solid and dashed lines indicate the 50 and 90$\%$ CI area of the event, respectively. It can be seen that the 90\% probability region of the event is covered almost entirely by DECam with $\sim 15$ pointings.}
    \label{fig:184g_i}
\end{figure}

It is worth noting that, at least in our GW population, the volumetric merger rate of compact objects in the NS mass regime decreases with mass. Hence, the most massive systems are rarer than lower-mass objects. For this reason, and because of the larger volume probed by the O5 network, a larger fraction of GW detections are expected to promptly collapse in a BH
which diminishes the fraction of objects producing a SMNS or an HMNS (see the last columns of Table \ref{table bns remnant}).

\subsection{Sky tiling and overall detection rate}

\begin{table}[!ht]
    \begin{tabular}{c|c|c|c}
        \hline
        \hline
        Run & Binary type & \multicolumn{2}{c}{KN detectability} \\
        \hline
        & & $g$-band & $i$-band \\
        \hline
        & & (\%) & (\%) \\
        \hline
        \hline
        \multirow{2}{*}{O4} & BNS & 77 & 77 \\
                            & NSBH & 6 & 6 \\
        \hline
        \multirow{2}{*}{O5} & BNS & 69 & 66 \\
                            & NSBH & 3 & 3 \\
    \hline
    \end{tabular}
    \caption{KN detection efficiencies (out of the total GW detections) using our observing strategy for a fiducial depth of $g$ = 24.3 and $i$ = 23.9 and the fiducial EOS. Here O4 refers to simulation 2}
    \label{tab:kn_det}
\end{table}

For each simulated event, we use \texttt{gwemopt} to produce an observing plan, which determines whether our default observing strategy would cover the true location of the merger. Combined with the detectability results from the previous subsection, this provides an estimate of the percentage of events that may have a KN detected in the O4 and O5 observing runs.

For example, Figure \ref{fig:184g_i} shows the DECam tilings for an example BNS event from simulation 2 at 174 Mpc localized to 37 sq. deg. at 90\% CI. In O4, only 3$\%$ of the BNS events are expected to be so well-localized. We find that DECam is able to cover the 90\% probability region for this well-localized event almost entirely with only $\sim 17$ pointings. Our baseline case assumes the follow up is performed 1 day after the GW trigger.

\begin{figure*}[htpb!]
    \centering
    \includegraphics[width=\linewidth]{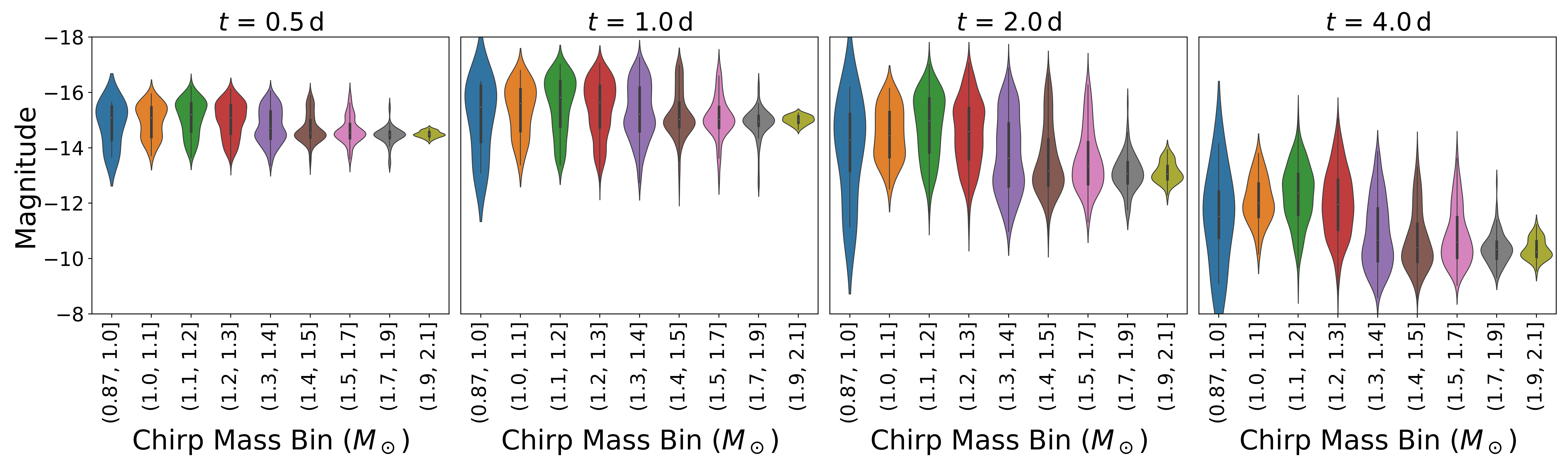}
    \includegraphics[width=\linewidth]{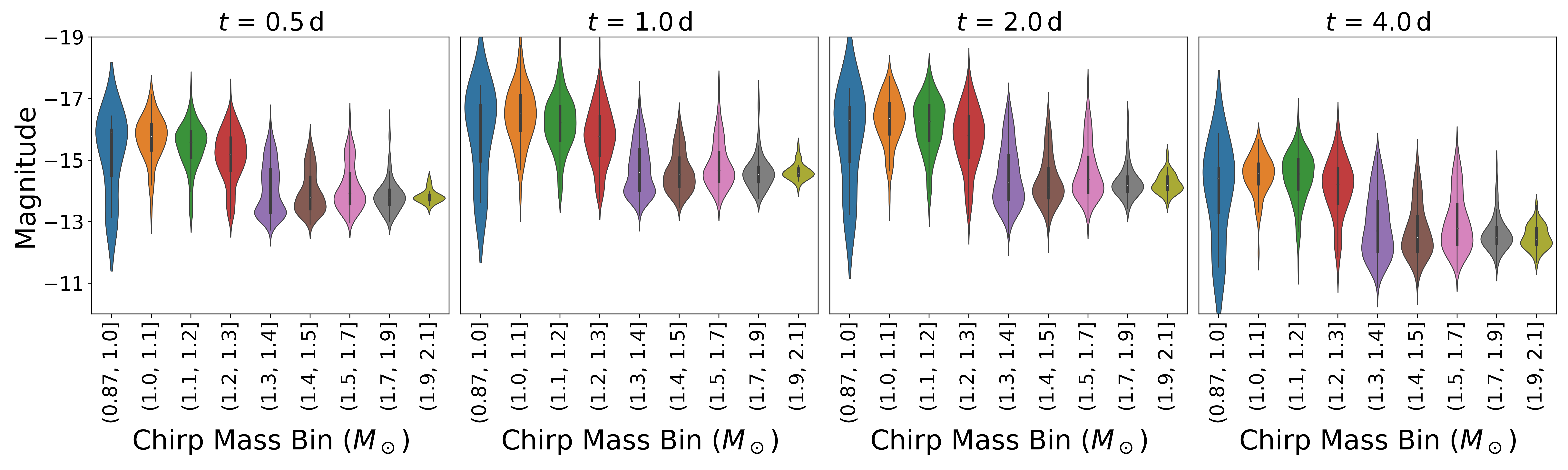}
    \caption{BNS kilonova magnitude distributions at 0.5,1,2 and 4 days as a function of chirp mass bins. The top panel shows the results in g-band, the bottom panel in i-band.}
    \label{fig:knmagbns}
\end{figure*}

\begin{figure*}[htpb!]
    \centering
    \includegraphics[width=\linewidth]{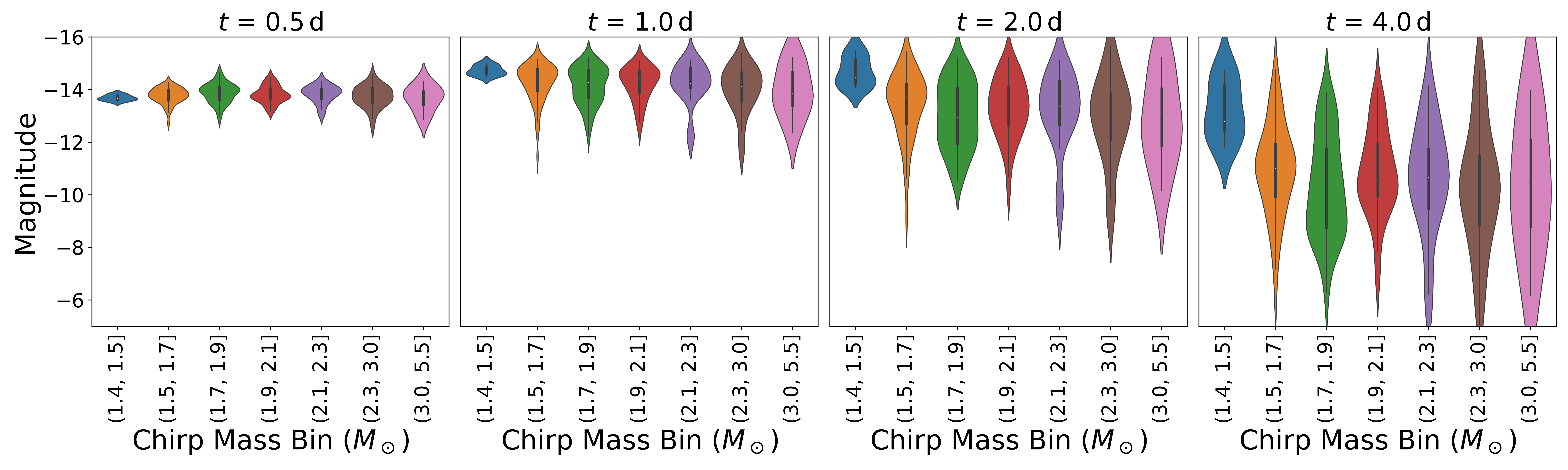}
    \includegraphics[width=\linewidth]{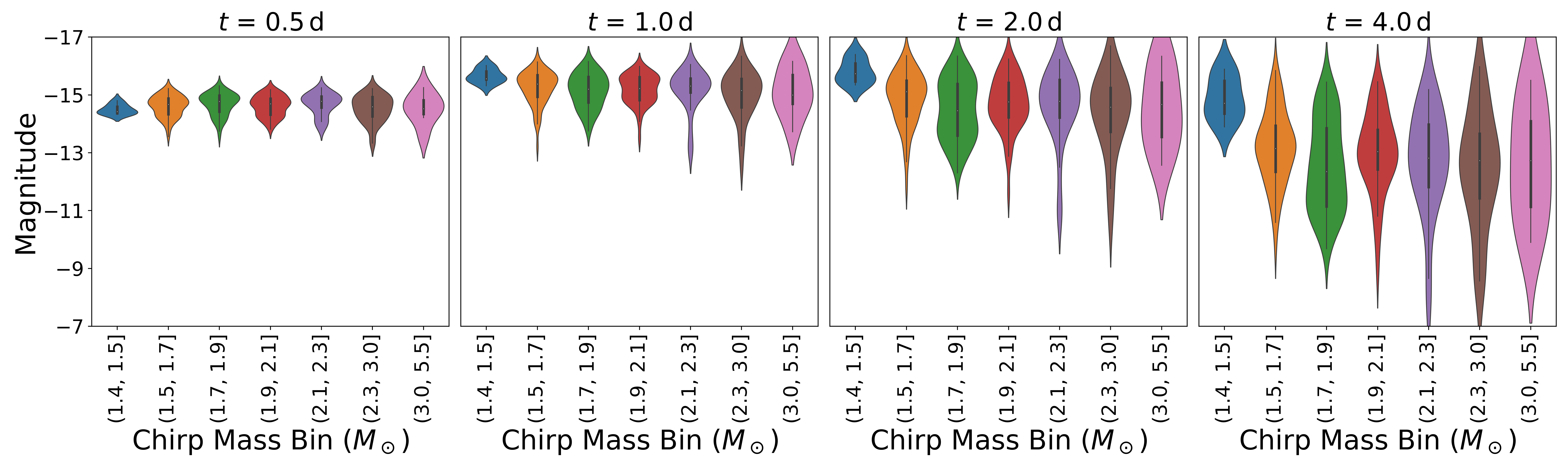}
    \caption{NSBH kilonova magnitude distributions at 0.5,1,2 and 4 days as a function of chirp mass bins. The top panel shows the results in g-band, the bottom panel in i-band.}
    \label{fig:knmagnsbh}
\end{figure*}

The results shown above only assume a single value of the depth for all events. The KN detection efficiency out of the total GW detection is shown in Table {\ref{tab:kn_det}}. In reality, we want to adjust the depth (and filters) based on the Moon phase and findings of this work for the KN population at different distances. For this reason, we provide the quintiles of our KN population at different distances and times in Appendix in Table \ref{tab:mags bns} for BNS events and \ref{tab:mags nsbh} for NSBH events, as well as the violin plots of KN brightness as a function of chirp mass and time for both BNS (Figure \ref{fig:knmagbns}) and NSBH (Figure \ref{fig:knmagnsbh}) so that these can be used to inform the depth of follow-up campaigns with DECam and other instruments. 
\begin{table}[!ht]
    \centering
    \begin{tabular}{c|c|c|c|c}
        \hline
        \hline
        Run & Binary type & \texttt{gwemopt} coverage  & \multicolumn{2}{c}{KN detectability} \\
        \hline
        & & & $g$-band & $i$-band \\
        & & (\%) & (\%) & (\%) \\
        \hline
        \hline
        \multirow{2}{*}{O4} & BNS & 32 & 97 & 88 \\
                            & NSBH & 29 & 99 & 100 \\
        \hline
        \multirow{2}{*}{O5} & BNS & 38 & 98 & 92 \\
                            & NSBH & 32 & 72 & 98 \\
    \hline
    \end{tabular}
    \caption{Detection efficiencies using \texttt{gwemopt} for different scenarios assuming the fiducial EOS and our fiducial observing strategy. The KN detectability columns shows the percentage of kilonovae covered by \texttt{gwemopt} that is detectable in $g$ or $i$-band given our fiducial depth. Results for the softer and stiffer EOSs are shown in Appendix \ref{sec:appendix1}.}
    \label{tab:detection_efficiency_fid}
\end{table}

For our case, we run \texttt{gwemopt} assuming these different depths of up to 24.3 and 23.9 in $g$ and $i$, respectively, and a maximum on-sky time of 4 hours. With these optimized settings, in O4 (O5), we expect to cover about 32\% (38\%) of all BNS KNe within 1 day of explosion and 29\% (32\%) of NSBH KNe within 2 days of explosion. Refer to Table \ref{tab:detection_efficiency_fid} in the Appendix, which summarizes the detection efficiency using \texttt{gwemopt} in the $g$ and $i$ bands across all scenarios. These percentages now reflect both the fact that the location of the KN is observable and covered by the scheduler and that the KN luminosity was above the flux limit. Considering the observability constraints that we place on airmass and moon distance, it is reasonable that about $\sim38\%$ of KNe from BNS events are covered by our DECam scheduler in O5. NSBH KNe result in a lower coverage probability because the depths needed to detect the EM counterpart are significantly larger, hence the longer exposure times required, resulting in smaller sky areas being covered in the fixed time window, despite the GW localizations of BNS and NSBH mergers appearing similar in our simulations. 

Next, we consider the total time required to follow up one event over multiple epochs to understand whether a down-selection of events is needed to use a reasonable amount of telescope time. For O4 the GW detection rates are low enough that a down-selection of events is unnecessary, so we consider the case of O5. For a fiducial cadence of 3 epochs (day 1, 2, and 4 post merger), we assess the feasibility of follow-up within realistic observing constraints, using the GW-MMADS time allocation of $\sim13$ hours per month for BNS/NSBH events. We restrict our analysis to events with chirp mass $\mathcal{M} < 3~M_\odot$, sky localization area $< 2000~\mathrm{deg}^2$, and at least $5\%$ of the sky localization probability covered by the simulated follow-up (using \texttt{gwemopt}). For BNS events satisfying these criteria, we find that one event requires a median of $\sim7$ hours of telescope time, with the $90^{\mathrm{th}}$ percentile requiring $\leq17.5$ hours for full three-epoch coverage. These events comprise approximately $23\%$ of the BNS events detectable in O5, corresponding to an expected $\sim2$--$17$ events per year. Given the 13-hour monthly observing allowance, this translates to the practical feasibility of following up $\sim1$ such BNS event per month. For NSBH systems, the estimated O5 detection rate is $\sim19$--$110~\mathrm{yr}^{-1}$. Applying the same selection criteria, we find that a typical NSBH event requires a median time of $\sim4.6$ hours for follow-up, with the $90^{\mathrm{th}}$ percentile requiring $\leq12$ hours. These represent approximately $8\%$ of the detectable NSBH population, or $\sim1.8$--$8.9$ events per year. Therefore, even for NSBH mergers, a limited but meaningful subset of events can be feasibly observed within the available time.

Next, we compute how many events we expect to detect in GW and also jointly with a KN, based on the GW detection rates and our KN detection efficiency. For DECam-like instruments, we estimate BNS (NSBH) KN detectability rate of $1-16~\mathrm{yr}^{-1}$ ($0-1~\mathrm{yr}^{-1}$) in O4 and $18-194~\mathrm{yr}^{-1}$ ($0-4~\mathrm{yr}^{-1}$) in O5 respectively. These ranges take into account the uncertainty on the latest rate estimates and on the EOS, with the rates contributing the most to the final uncertainty. These rates drop if corrected for our \texttt{gwemopt} observability efficiency, which can be calculated from Tables \ref{tab:detection_efficiency_soft} and \ref{tab:detection_efficiency_stiff} in Appendix \ref{sec:appendix1} as $\sim 0-5~\mathrm{yr}^{-1}$ ($0~\mathrm{yr}^{-1}$) and $\sim 7-76~\mathrm{yr}^{-1}$ ($\sim 0-1~\mathrm{yr}^{-1}$) multimessenger BNS (NSBH) detections in O4 and O5. If we consider the non-detection of BNS mergers in O4 until the end of O4b (see Table \ref{tab:detection_rate_O4b}), the BNS  KN detectability rate drops  to $0-5~\mathrm{yr}^{-1}$ in O4 and $5 - 55~\mathrm{yr}^{-1}$ in O5;  when corrected for our \texttt{gwemopt} observability efficiency, the rate drops to $0 - 2~\mathrm{yr}^{-1}$ in O4 and $2 - 21~\mathrm{yr}^{-1}$ in O5.
 \begin{table*}[!htpb]
    \centering
    \begin{tabular}{c|c|c|c|c}
        \hline
        \hline
        Run & Binary type & GW detection rate & \multicolumn{2}{c}{KN detectability} \\
        \hline
        & & & $g$-band & $i$-band \\
        & & ($\mathrm{yr}^{-1}$) & ($\mathrm{yr}^{-1}$) & ($\mathrm{yr}^{-1}$) \\
        \hline
        \hline
        \multirow{2}{*}{O4} & BNS & 0.60 -- 5.7 & 0.40 -- 4.6 & 0.40 -- 4.6 \\
        \cline{2-5}
        & NSBH & 1.7 -- 9.8 & 0.11 -- 0.66 & 0.10 -- 0.70 \\
        \hline
        
        \multirow{2}{*}{O5} & BNS & 8.6 -- 76 & 5.3 -- 55 & 5.1 -- 53 \\
        \cline{2-5}
        & NSBH & 19 -- 110 & 0.45 -- 3.6 & 0.54 -- 4.3 \\
        \hline
    \end{tabular}
    \caption{Expected BNS and NSBH detection rates from our LVK O4 and O5 simulations and the associated multimessenger KN detection rates in g and i bands, assuming different EOSs. The BNS rate caluclation assumes that there were no BNS detections in O4 until the end of 2024.}
    \label{tab:detection_rate_O4b}
\end{table*}
We note that even though the O4 GW BNS merger rates we find may appear large given the non-detection of high-confidence BNS candidates in O4, our detection rate (as also in other works such as \citealt{Kiendrebeogo_2023}) is based on an SNR cut that does not correspond exactly to the O4 search pipelines selection criteria, based on False Alarm Rate (FAR) instead. Therefore, our selection may include events that would be considered sub-threshold in the actual O4 searches. However, the expected number of multimessenger sources is likely to be closer to what we will actually find in O4, modulo the uncertain rates that may push us towards the lower limits we report, given that those sources are typically the better localized and more nearby, and will have the largest SNR and lower FAR of the entire BNS population. In our simulations release, we include detector and network SNRs, so our findings can easily be rescaled in light of the new findings from the O4 search pipelines and the relation between their FARs and SNRs, as well as in light of updated rates.

Lastly, we point to the fact that different EOS will predict slightly different KN emission and therefore detection rates (e.g. \citealt{1999A&A...341..499R,Coughlin_2018,Zhao_2023}). We find that the detection rate is only mildly dependent on the EOS, so that the uncertainty on our detection rate is dominated by the underlying compact object volumetric merger rate uncertainty rather than by changes in the EOS.

\section{Discussion}\label{section:Discussion}

Although we have assumed the same relations to map BNS properties into ejecta masses regardless of the fate of the central remnant, some differences may be expected for different remnants~\citep{Kawaguchi_2020}, especially when a stable NS is formed. The major difference is expected to be in the wind/outflows mass ejecta rather than in the dynamical ejecta. However, this accounts for $<2\%$ of all BNS systems considered in both O4 and O5, and it is therefore only expected to be a minor perturbation for our results.

Next, we compare our work with previous efforts on GW localizations and KN detectability. Compared to \citet{2018LRR....21....3A} we consider more up-to-date detector sensitivities and simulate KN lightcurves. The main differences from the studies in \citet{colombo22,colombo24,shah24} are that we assume a mass distribution driven by GW observations and provide the depth needed to detect a significant fraction of BNS and NSBH KN. 

In \citet{Kiendrebeogo_2023} the aim is to understand how multimessenger observations will help in estimating the Hubble constant with a focus on optimizing follow-up using Zwicky Transient Facility (ZTF) and Rubin Observatory. Simulated follow-up with ZTF is modeled in $g,r,i$ bands whereas for Rubin Observatory simulated follow-up are modeled in $u,g,r,i,z,y$ bands. Follow-up decisions for going deep or wide are made based on GW candidate properties such as FAR and presence or absence of neutron star remnant. An event satisfying the ``go-deep'' criteria merits triggering a target of opportunity observation for 3 nights in $g$ and $r$ bands for 300s exposure, whereas an event satisfying ``go-wide'' criteria would prompt reweighing the ZTF survey field to observe the event for 5 nights of 30s exposures. Instead, our approach of simulating GW KNe directly from a population model marginalized over GWTC-3 uncertainties and using the predicted lightcurves to infer their expected magnitude allows the depth and likelihood of detection to be continuously determined by physical parameters such as binary chirp mass, EOS, viewing angle, and distance rather than applying pre-assigned depth tiers. Moreover, by marginalizing over the GW population uncertainty we get a different and more numerous mass gap NSBH sample as the population hyperparameters used in \citet{Kiendrebeogo_2023} predict an almost empty mass gap.

We also add to the work presented in \citet{bom24}, which focuses on DECam follow up in O4, by using more updated gravitational wave simulations, physically motivated KN light curves, as well as the addition of O5 forecasts. \citet{bom24} design DECam strategies based on blue and red isotropic kilonova models from \citet{Kasen_2017}, and optimize for cadence and depth to improve detection efficiency across $g,r,i,z$ bands. On the other hand, in this work we capture population level uncertainty from the GW sources and use different kilonova models with geometries motivated by numerical relativity simulations and with properties informed by the GW sources.

\citet{Biscoveanu_2022} also explore multimessenger prospects for NSBH mergers based on the findings from GWTC-3. A major difference from that work is that they impose an empty lower mass gap, while we allow it to be populated with mass gap NSBHs, which is reasonable based on the detection of GW230529 \citep{230529_LVK}. This is expected to produce a larger number of NSBHs giving rise to an EM counterpart in our work. On the other hand, given the lack of NSBH detections with BH mass $>20~M_\odot$, their population of NSBH binaries has a BH mass cut off at $>20~M_\odot$. Because the GW population assumed in our analysis does not distinguish between compact object binary types and can therefore include higher mass BHs following \citet{Farah_2022}, our BH cutoff is effectively at $60~M_\odot$ for NSBH binaries. Although this choice may reduce the fractional number of multimessenger NSBH sources compared to a lower BH mass cutoff, the pairing function would rarely allow for the presence of extreme mass ratio pairings between a $>20~M_\odot$ BH and a NS (it can be seen from Figure \ref{fig:m1m2} that $m_{1}$ in our NSBH population has already tailed off at $13~M_\odot$). In our simulations, as we go from softer to stiffer EOS, we find that $\sim 6-8\%$ of NSBHs in O4 are expected to produce a KN of which 0.3$^{+0.3}_{-0.2}$ - 0.3$^{+0.4}_{-0.2}$ per year would be detectable in O4 given our fiducial depth. In O5, we find that $\sim 6-9\%$ of NSBH events are expected to produce a KN depending on the EOS and 1.2$^{+1.4}_{-0.8}$ - 2.0$^{+2.3}_{-1.3}$ per year would be detectable given our fiducial depth. As a result, our prospects for detecting multimessenger NSBH events are promising.

Comparing the rate estimates from similar works, \citet{Petrov_2022} reports BNS detection rates of $9$-$88~\mathrm{yr}^{-1}$ in O4 and $47$-$478~\mathrm{yr}^{-1}$ in O5, and NSBH rates of $34$--$147~\mathrm{yr}^{-1}$ in O4 and $180$--$720~\mathrm{yr}^{-1}$ in O5. \citet{colombo22} provides O4 rate estimates of $2$-$20~\mathrm{yr}^{-1}$ for BNS and \citet{colombo24} reports a rate of $6$-$39~\mathrm{yr}^{-1}$ for NSBH, and O5 rate estimates of $56$--$292~\mathrm{yr}^{-1}$ for NSBH. Our estimates are lower, particularly for O4, as we incorporate the non-detection of BNS mergers through the end of O4b, providing a more stringent and up-to-date constraint based on the LVK public alerts to date.

Our findings reveal strong dependencies of KN detection efficiency on the binaries mass (and spin for NSBHs) parameters. This is not surprising as our mapping between binary parameters and ejecta mass explicitly includes these quantities. However, this also indicates that a significant EM selection function will be at play and should be taken into account when multimessenger studies of populations of BNSs and NSBHs, such as EOS and standard siren measurements, will become possible. 

To conclude, if spins and mass ratios were to be released in public GW alerts in addition to chirp mass, this would be an invaluable tool for the astronomical community to further inform their follow-up \citep{Margalit_2019} and, for example, prioritize NSBH mergers with larger effective spin. A caveat is that we have considered detecting any KN that is brighter in at least one band than the magnitude limit considered in our baseline observing strategy. In reality, one would typically require at least two detections of the transient in order to discriminate between moving objects (a major contaminant in transient searches) and static variables, while also identifying transients that most resemble the colors and evolution of a KN. Various tools and methods can be used to rapidly identify them, even with a few data points (e.g., \citealt{ofek}).

\section{Conclusion}\label{section:Conclusion}

In this work, we describe an end-to-end simulation of GW merger events for the fourth and fifth LIGO/Virgo/KAGRA observing runs to optimize follow-up strategies with our GW-MMADS DECam Survey program. We focus on mergers containing at least one neutron star. We study both the BNS and NSBH parameters and their effect on kilonova production and detectability, finally optimizing our observing strategy based on the expected KN population. 
The implications of this simulation study can be summarized as follows:\\
(i) Including Virgo, even at O3 sensitivity, the O4 detector network results in larger detection rates of well-localized events at distances $<$100 Mpc. Approximately $15\%$ of the BNS and NSBH events are likely to result in a localization that is $<500$ sq.\ deg.\ in O4, which in O5, increases to $26\%$ for BNS and $28\%$ for NSBH mergers respectively. More than $10\%$ of the neutron star merger events will have an area localized to $<100~\mathrm{deg}^2$ in O5.\\
(ii) For our baseline strategy (without considering the \texttt{gwemopt} coverage), we find that in O5, out of all GW detections, KNe are detectable for $\sim 69\% (66\%)$ of BNS mergers and $\sim 3\% (3\%)$ NSBH mergers, at a depth of 24.3 and 23.9 mag in $g$ and $i$ respectively. These depths can be optimized for a specific follow-up given an event's distance; we provide tables with quantiles of the magnitudes for our expected KN population to allow for that.\\
(iii) The majority of BNS sources detected in GWs as well as those that have a KN detection are expected to produce a HMNS remnant, while a significant fraction of the remaining detections are likely to undergo prompt collapse to a BH.\\ 
(iv) We show the dependence of KN detectability on GW parameters, providing a realistic population of GW detections. In particular, we show the importance of releasing even tentative chirp masses and mass ratios to the EM follow-up community to inform their triggering criteria and follow-up strategy. The release of binned chirp masses in O4 is an important step in that direction. We provide the expected kilonova magnitude as a function of chirp mass bins to further inform the GW-EM follow-up efforts.\\
(vi) 90th percentile of multimessenger BNS detection rate is expected to be below a chirp mass of $<1.7~M_\odot$ and at distances less than 1140 Mpc. The KN detection efficiency out of the GW detections is also higher for mass ratios $<0.8$ compared to more symmetric BNSs. For NSBH mergers, 90th percentile of the detectable KNe have a chirp mass $<2.4~M_\odot$, mass ratios of $[0.2,0.6]$, and effective spin $-0.2<\chi_{\rm eff}<0.7$, i.e., strongly favoring co-aligned BH spins with respect to the angular momentum of the binary. \\
(vii) For DECam-like instruments, considering the non-detection of BNS mergers in O4 up until the end of O4b, we estimate a KN detectability rate of $\sim 0 - 5~\mathrm{yr}^{-1}$ from BNS mergers and $0 - 1~\mathrm{yr}^{-1}$ from NSBH mergers in O4. In O5, our estimated KN detectability rate is $\sim 5 - 55~\mathrm{yr}^{-1}$ from BNS mergers and $\sim 0 - 4~\mathrm{yr}^{-1}$ from NSBH mergers. Taking into account our typical \texttt{gwemopt} coverage, this leads to $\sim 0-2$ ($\sim 2-21)~\mathrm{yr}^{-1}$ multimessenger BNS detection and probably no ($\sim 0-1~\mathrm{yr}^{-1}$) multimessenger NSBH detection in O4 (O5). Our expectation is that there is a significant probability of detecting an EM counterpart to an NSBH merger in O5.

These findings can be useful to inform a variety of follow-up efforts, especially those that will soon be carried out by the Rubin Observatory \citep{RubinToO}.

\nopagebreak
\section*{Acknowledgements}
\noindent Antonella Palmese acknowledges support for this work by NSF Grant No. 2308193. Mattia Bulla acknowledges
the Department of Physics and Earth Science of the
University of Ferrara for the financial support through
the FIRD 2024 grant. Tim Dietrich acknowledges funding from the European Union (ERC, SMArt, 101076369). Views and opinions expressed are those of the authors only and do not necessarily reflect those of the European Union or the European Research Council. Neither the European Union nor the granting authority can be held responsible for them.
Brendan O'Connor is supported by the McWilliams Postdoctoral Fellowship at Carnegie Mellon University. This research used resources of the National Energy Research
Scientific Computing Center, a DOE Office of Science User Facility
supported by the Office of Science of the U.S. Department of Energy
under Contract No. DE-AC02-05CH11231 using NERSC award
HEP-ERCAP0029208 and HEP-ERCAP0022871. This work used resources on the Vera Cluster at the Pittsburgh Supercomputing Center. We thank T.J. Olesky and the PSC staff for help with setting up our software on the Vera Cluster.\\
\appendix
\section{Kilonova magnitude table}\label{sec:appendixknmag}

We show the expected 50 and 90 percentile quantiles of KN magnitude distributions from 0.5 days to 5 days following BNS (Table \ref{tab:mags bns}) and NSBH (Table \ref{tab:mags nsbh}) at O4 and O5 sensitivities.
\begin{table*}[!ht]
\centering
\begin{tabular}{ c | c | c || c | c | c | c | c | c || c | c | c | c | c | c }
\hline
\hline
Distance & Observing run & band & \multicolumn{6}{c|}{mag$_{50}$} & \multicolumn{6}{c}{mag$_{90}$}\\
\hline
(Mpc) & & & 12h & 1d & 2d & 3d & 4d & 5d & 12h & 1d & 2d & 3d & 4d & 5d\\
\hline \hline
\multirow{4}{*}{$<$ 100} & O4 & $g$ & 19.0 & 18.4 & 19.5 & 21.0 & 22.3 & 23.2 & 20.2 & 20.3 & 21.4 & 22.6 & 23.6 & 24.5\\

& &$i$ & 19.0 & 18.2 & 19.3 & 20.2 & 21.0 & 21.7 & 20.0 & 19.3 & 19.9 & 21.1 & 21.9 & 22.7\\
 
& O5 & $g$ & 19.8 & 19.8 & 20.1 & 21.1 & 22.0 & 22.8 & 20.2 & 20.3 & 21.4 & 22.8 & 23.9 & 24.9
\\

&  & $i$ & 19.5 & 18.7 & 19.1 & 19.7 & 20.3 & 20.8 & 20.3 & 19.6 & 19.9 & 21.1 & 22.1 & 22.9\\
\hline

\multirow{4}{*}{100-150} & O4 & $g$ & 20.3 & 19.8 & 20.7 & 22.1 & 23.3 & 24.2 & 21.2 & 21.2 & 21.8 & 23.2 & 24.4 & 25.4\\

& & $i$ & 20.5 & 19.7 & 20.4 & 21.2 & 22.0 & 22.7 & 21.2 & 20.6 & 21.0 & 22.3 & 23.2 & 23.9\\
 
& O5 &$g$ & 19.8 & 19.1 & 19.9 & 21.3 & 22.6 & 23.6 & 20.6 & 20.4 & 21.7 & 22.7 & 23.6 & 24.5\\
 
&  & $i$ & 20.3 & 19.5 & 20.3 & 21.0 & 21.6 & 22.3 & 20.8 & 20.4 & 20.7 & 21.8 & 22.7 & 23.4\\
\hline

\multirow{4}{*}{150-200} & O4 & $g$ & 20.7 & 20.1 & 21.1 & 22.4 & 23.7 & 24.7 & 21.7 & 21.4 & 22.2 & 23.5 & 24.9 & 25.9\\

 & &$i$ & 20.9 & 20.1 & 21.1 & 22.0 & 22.8 & 23.5 & 21.5 & 20.8 & 21.6 & 22.9 & 23.8 & 24.6\\
 
 & O5 & $g$ & 21.0 & 20.3 & 21.4 & 22.7 & 24.1 & 25.0 & 21.9 & 21.8 & 23.0 & 24.2 & 25.0 & 25.8\\
 
 &  & $i$ & 21.1 & 20.3 & 21.3 & 22.2 & 23.0 & 23.8 & 21.8 & 21.4 & 21.6 & 22.8 & 23.7 & 24.4\\
\hline

\multirow{4}{*}{200-250} & O4 & $g$ & 21.4 & 20.7 & 21.6 & 23.0 & 24.2 & 25.1 & 22.1 & 21.8 & 22.8 & 24.2 & 25.3 & 26.3\\

& & $i$ & 21.5 & 20.7 & 21.4 & 22.2 & 23.1 & 23.8 & 22.0 & 21.3 & 22.1 & 23.3 & 24.2 & 25.0\\
 
& O5 & $g$ & 21.6 & 21.0 & 21.9 & 23.1 & 24.4 & 25.1 & 22.2 & 22.2 & 23.0 & 24.4 & 25.4 & 26.3\\
 
&  & $i$ & 21.7 & 20.9 & 21.3 & 21.9 & 22.6 & 23.2 & 22.5 & 22.0 & 22.0 & 23.2 & 24.1 & 24.8\\
\hline

\multirow{4}{*}{250-300} & O4 & $g$ & 21.8 & 21.0 & 22.0 & 23.4 & 24.6 & 25.5 & 22.4 & 22.0 & 23.1 & 24.5 & 25.7 & 26.8\\

& & $i$ & 21.9 & 21.0 & 21.7 & 22.6 & 23.4 & 24.0 & 22.5 & 21.7 & 22.4 & 23.7 & 24.7 & 25.4\\
 
& O5 & $g$ & 21.9 & 21.3 & 22.2 & 23.4 & 24.6 & 25.5 & 22.5 & 22.4 & 23.2 & 24.5 & 25.7 & 26.7\\
 
&  & $i$ & 22.1 & 21.2 & 21.8 & 22.5 & 23.2 & 23.8 & 22.7 & 22.2 & 22.4 & 23.6 & 24.5 & 25.3\\
\hline

\multirow{4}{*}{300-500} & O4 & $g$ & 22.5 & 21.7 & 22.5 & 23.9 & 25.1 & 26.1 & 23.3 & 22.9 & 23.8 & 25.1 & 26.4 & 27.4\\

& & $i$ & 22.6 & 21.7 & 22.4 & 23.3 & 24.1 & 24.8 & 23.2 & 22.4 & 23.2 & 24.4 & 25.3 & 26.0\\
 
& O5 & $g$ & 22.7 & 21.9 & 22.8 & 24.2 & 25.3 & 26.2 & 23.6 & 23.3 & 24.0 & 25.3 & 26.5 & 27.5\\
 
&  & $i$ & 22.9 & 22.0 & 22.5 & 23.3 & 24.0 & 24.7 & 23.6 & 22.9 & 23.3 & 24.5 & 25.5 & 26.2\\
\hline

\end{tabular}
\caption{50th and 90th quantiles of the KN magnitude distributions at 0.5,1,2,3,4 and 5 days for the BNS mergers at O4 and O5 sensitivities.}
\label{tab:mags bns}
\end{table*}

\begin{table*}[!ht]
\begin{center}
\begin{tabular}{ c | c | c || c | c | c | c | c | c || c | c | c | c | c | c }
\hline
\hline
Distance & Observing run & band & \multicolumn{6}{c|}{mag$_{50}$} & \multicolumn{6}{c}{mag$_{90}$}\\
\hline
(Mpc) & & & 12h & 1d & 2d & 3d & 4d & 5d & 12h & 1d & 2d & 3d & 4d & 5d\\
\hline \hline

\multirow{4}{*}{$<$ 200} & O4 & $g$ & 22.6 & 21.6 & 21.6 & 21.9 & 22.8 & 23.6 & 23.1 & 22.2 & 22.2 & 22.7 & 23.6 & 24.4\\

 & &$i$ & 20.7 & 20.1 & 20.0 & 20.4 & 20.9 & 21.4 & 21.9 & 21.1 & 21.0 & 21.2 & 21.7 & 22.3\\
 
& O5 & $g$ & 22.3 & 21.4 & 21.4 & 21.8 & 22.6 & 23.5 & 22.6 & 21.7 & 21.8 & 22.4 & 23.1 & 23.8\\

&  & $i$ & 20.6 & 19.9 & 19.9 & 20.2 & 20.5 & 21.0 & 21.1 & 20.5 & 20.4 & 20.5 & 20.9 & 21.4\\
\hline

\multirow{4}{*}{200-300} & O4 & $g$ & 23.6 & 22.5 & 22.6 & 23.1 & 23.8 & 24.6 & 24.0 & 22.9 & 23.0 & 23.6 & 24.3 & 25.1\\

& &$i$ & 21.7 & 21.3 & 21.1 & 21.4 & 21.8 & 22.2 & 22.3 & 21.7 & 21.6 & 22.1 & 22.5 & 23.2\\
 
 & O5 & $g$ & 24.0 & 22.9 & 23.0 & 23.5 & 24.2 & 25.0 & 24.3 & 23.3 & 23.3 & 23.9 & 24.7 & 25.5\\
 
 &  & $i$ & 22.1 & 21.4 & 21.4 & 21.8 & 22.1 & 22.6 & 23.4 & 22.5 & 22.4 & 22.4 & 22.8 & 23.4\\
\hline

\multirow{4}{*}{300-400} & O4 & $g$ & 24.2 & 23.0 & 23.1 & 23.6 & 24.3 & 25.0 & 24.5 & 23.3 & 23.4 & 23.9 & 24.6 & 25.4\\

& &$i$ & 22.4 & 21.7 & 21.7 & 22.0 & 22.4 & 22.8 & 22.9 & 22.2 & 22.2 & 22.5 & 23.0 & 23.5\\
 
 & O5 & $g$ & 24.5 & 23.5 & 23.5 & 24.0 & 24.7 & 25.5 & 25.0 & 23.9 & 23.8 & 24.3 & 25.1 & 26.0\\
 
 &  & $i$ & 22.8 & 22.1 & 22.0 & 22.4 & 22.8 & 23.3 & 23.7 & 23.0 & 22.9 & 23.3 & 23.8 & 24.3\\
\hline

\multirow{4}{*}{400-500} & O4 & $g$ & 25.0 & 23.8 & 24.0 & 24.5 & 25.1 & 25.8 & 25.4 & 24.0 & 24.1 & 24.6 & 25.3 & 26.1\\

 & & $i$ & 23.0 & 22.4 & 22.3 & 22.6 & 23.0 & 23.4 & 23.2 & 22.8 & 22.8 & 23.2 & 23.6 & 24.2\\
 
 & O5 & $g$ & 24.8 & 23.8 & 23.8 & 24.2 & 24.9 & 25.6 & 25.3 & 24.0 & 24.1 & 24.6 & 25.3 & 26.0\\
 
 &  & $i$ & 23.0 & 22.4 & 22.3 & 22.6 & 23.0 & 23.4 & 23.6 & 22.9 & 22.9 & 23.2 & 23.8 & 24.4\\
\hline

\multirow{4}{*}{500-600} & O4 & $g$ & 25.4 & 24.2 & 24.2 & 24.6 & 25.2 & 25.9 & 25.8 & 24.4 & 24.5 & 25.0 & 25.7 & 26.5\\

& & $i$ & 23.5 & 23.0 & 22.9 & 23.2 & 23.4 & 23.8 & 24.3 & 23.4 & 23.4 & 23.8 & 24.3 & 25.0\\
 
& O5 & $g$ & 25.4 & 24.3 & 24.3 & 24.8 & 25.4 & 26.1 & 25.7 & 24.5 & 24.5 & 25.0 & 25.7 & 26.5\\
 
&  & $i$ & 23.5 & 22.8 & 22.8 & 23.1 & 23.4 & 23.8 & 24.3 & 23.4 & 23.4 & 23.7 & 24.1 & 24.7\\
\hline

\multirow{4}{*}{600-700} & O4 & $g$ & 25.5 & 24.4 & 24.4 & 24.9 & 25.5 & 26.2 & 25.9 & 24.6 & 24.7 & 25.2 & 25.9 & 26.6\\

& & $i$ & 23.6 & 23.0 & 22.9 & 23.3 & 23.7 & 24.2 & 24.1 & 23.3 & 23.3 & 23.7 & 24.2 & 24.8\\
 
& O5 & $g$ & 25.6 & 24.4 & 24.5 & 24.9 & 25.5 & 26.1 & 25.8 & 24.7 & 24.6 & 25.1 & 25.8 & 26.6\\
 
&  & $i$ & 23.8 & 23.1 & 23.0 & 23.3 & 23.5 & 23.9 & 24.4 & 23.5 & 23.4 & 23.8 & 24.3 & 25.0\\
\hline

\end{tabular}
\end{center}
\caption{50th and 90th quantiles of the KN magnitude distributions at 0.5,1,2,3,4 and 5 days for the NSBH mergers at O4 and O5 sensitivities.}
\label{tab:mags nsbh}
\end{table*}

\section{Detection efficiency estimate}\label{sec:appendix1}

We calculate the \texttt{gwemopt} coverage and the associated KN detectability for the simulated events also assuming our softer and stiffer EOSs. The results are shown below in Table
\ref{tab:detection_efficiency_soft} and Table \ref{tab:detection_efficiency_stiff}.
\begin{table}[!ht]
    \centering
    \begin{tabular}{c|c|c|c|c}
        \hline
        \hline
        Run & Binary type & \texttt{gwemopt} coverage  & \multicolumn{2}{c}{KN detectability} \\
        \hline
        & & & $g$-band & $i$-band \\
        & & (\%) & (\%) & (\%) \\
        \hline
        \hline
        \multirow{2}{*}{O4} & BNS & 30 & 98 & 95 \\
                            & NSBH & 26 & 90 & 99 \\
        \hline
        \multirow{2}{*}{O5} & BNS & 37 & 99 & 88 \\
                            & NSBH & 28 & 65 & 93 \\
    \hline
    \end{tabular}
    \caption{Detection efficiencies using our observing strategy for different scenarios and binaries using the softer EOS.}
    \label{tab:detection_efficiency_soft}
\end{table}

\begin{table}[!ht]
    \centering
    \begin{tabular}{c|c|c|c|c}
        \hline
        \hline
        Run & Binary type & \texttt{gwemopt} coverage & \multicolumn{2}{c}{Detectability} \\
        \hline
        & & & $g$-band & $i$-band \\
        & & (\%) & (\%) & (\%) \\
        \hline
        \hline
        \multirow{2}{*}{O4} & BNS & 33 & 99 & 99 \\
                            & NSBH & 31 & 98 & 100 \\
        \hline
        \multirow{2}{*}{O5} 
        & BNS & 39 & 98 & 95 \\
                            & NSBH & 33 & 72 & 100 \\
    \hline
    \end{tabular}
    \caption{Detection efficiencies using our observing strategy for different scenarios and binaries using the stiffer EOS.}
    \label{tab:detection_efficiency_stiff}
\end{table}

\bibliographystyle{yahapj}
\bibliography{references1}

\end{document}